\documentclass[]{aa} 
\usepackage[varg]{txfonts}
\usepackage{graphicx,color}
\usepackage{dcolumn}
\usepackage{bm}
\usepackage{lineno, soul, ulem} 
\usepackage{hyperref,url}
\usepackage{xcolor}
\usepackage{array}
\usepackage{comment}

\usepackage{natbib}
\bibpunct{(}{)}{;}{a}{}{,} 

\newcommand\nustar{\textit{NuSTAR}}

\newcommand\xmm{\textit{XMM--Newton}}
\newcommand\fermi{\textit{Fermi}}

\newcommand{\lsim}{\mathrel{\mathop{\kern 0pt \rlap
  {\raise.2ex\hbox{$<$}}}
  \lower.9ex\hbox{\kern-.190em $\sim$}}}
\newcommand{\gsim}{\mathrel{\mathop{\kern 0pt \rlap
  {\raise.2ex\hbox{$>$}}}
  \lower.9ex\hbox{\kern-.190em $\sim$}}}

 \hypersetup{
    colorlinks=true,
    linkcolor=violet,
    filecolor=purple,      
    urlcolor=blue,
    citecolor=violet,
    pdftitle={Geminga's pulsar halo: an X-ray view}
    }
    
\def\amin{\ifmmode^{\prime}\else$^{\prime}$\fi}
\def\asec{\ifmmode^{\prime\prime}\else$^{\prime\prime}$\fi}

\def\simgt{\lower.5ex\hbox{$\; \buildrel > \over \sim \;$}}
\def\simlt{\lower.5ex\hbox{$\; \buildrel < \over \sim \;$}}

\def \aap  {A\&A}

\def \apjs  {ApJS}
\def \apjl  {ApJL}

\begin{document}

\title{Geminga's pulsar halo: an X-ray view}

\author{Silvia Manconi\inst{\ref{inst1}}\thanks{Corresponding author. Email: \texttt{manconi@lapth.cnrs.fr}} \and Jooyun Woo\inst{\ref{inst2}} \and Ruo-Yu Shang\inst{\ref{inst3}}  \and Roman Krivonos\inst{\ref{inst4}} \and Claudia Tang\inst{\ref{inst3}} \and Mattia Di Mauro\inst{\ref{inst5}} \and Fiorenza~Donato\inst{\ref{inst5},\ref{inst6}} \and 
Kaya Mori\inst{\ref{inst2}} \and Charles J. Hailey\inst{\ref{inst2}}}

\institute{Laboratoire d'Annecy-le-Vieux de
Physique Théorique (LAPTh), CNRS, USMB, F-74940 Annecy, France \label{inst1} 
\and 
Columbia Astrophysics Laboratory, 550 West 120th Street, New York, NY 10027, USA\label{inst2}
\and
Department of Physics and Astronomy, Barnard College, Columbia University, NY 10027, USA\label{inst3}
\and
Space Research Institute (IKI), 84/32 Profsouznaya str., Moscow 117997, Russian Federation\label{inst4}
\and
Istituto Nazionale di Fisica Nucleare, via P. Giuria, 1, 10125 Torino, Italy
\label{inst5}
\and
Department of Physics, University of Torino, via P. Giuria, 1, 10125 Torino, Italy
\label{inst6}
}

   \date{LAPTH--012/24}

\abstract{Geminga is the first pulsar around which a remarkable gamma-ray halo extending over a few degrees was discovered at TeV energies by MILAGRO, HAWC and later by H.E.S.S., and by \fermi{}-LAT in the GeV band.  
More middle-aged pulsars have exhibited gamma-ray halos, and they are now recognized as an emerging class of Galactic gamma-ray sources.
The emission appears in the late evolution stage of pulsars, and is most plausibly explained by inverse Compton scattering of CMB and interstellar photons by relativistic electrons and positrons escaping from the  pulsar wind nebulae.
These observations pose a number of theoretical challenges, particularly the origin of the inferred significantly lower effective diffusion coefficients around the pulsar when compared to typical  Galactic values. 
Tackling these questions requires constraining the ambient magnetic field properties, which can be achieved through X-ray observations. 
If the gamma-ray halos originate from a distribution of highly energetic electrons, synchrotron losses in the ambient magnetic fields of the same particles are expected to produce a diffuse X-ray emission with a similar spatial extension.\\
We present the most comprehensive X-ray study of the Geminga pulsar halo to date, utilising archival data from \xmm{} and \nustar{}. Our  X-ray analysis covers a broad bandwidth ($0.5\rm{-}79$~keV) and large field of view ($\theta\sim4^\circ$) for the first time. This is achieved by accurately measuring the background over the entire field of view, and taking into account both focused and stray-light X-ray photons from the pulsar halo with \nustar{}. We find no significant  emission and set robust constraints on the X-ray halo flux.  These are  translated to stringent constraints on the ambient magnetic field strength and the diffusion coefficient by using a physical model 
considering particle injection, diffusion and cooling over the pulsar’s lifetime, which is tuned by fitting multi-wavelength data.
Our novel methodology for modelling and searching for synchrotron X-ray halos can be applied to other  pulsar halo candidates.}

\maketitle

\bigskip

\section{Introduction}
Gamma-ray halos are emerging as a general characteristic of middle-aged (few 100 kyrs old) Galactic pulsars \citep{Lopez-Coto:2022igd,Liu:2022hqf,Fang:2022fof}. 
The archetypal gamma-ray halo is the few-degree emission observed around Geminga at multi-TeV  \citep{2009ApJ...700L.127A,HAWC:2017kbo,Guo:2021vuy,HESS:2023sbf} and at about tens of GeV \citep{DiMauro:2019yvh} gamma-ray energies.
Extended emissions consistent with being pulsar halos \footnote{Given they were discovered very recently, and their properties and differences with respect to the pulsar wind nebulae still under investigation, their nomenclature in the literature is varied, and includes their definition as 'TeV halos', 'inverse Compton halos', or just 'pulsar halos'. In this paper we adopt the definition of 'Geminga pulsar halo', which includes the emission observed at GeV to TeV energies, as well as the putative synchrotron emission at the keV energies we search for.}   have been observed also around a few other pulsars, such as Monogem \citep{HAWC:2017kbo} and PSR J0622+3749 \citep{LHAASO:2021crt}. 
The observed non-thermal gamma-ray emission extends  at least a few tens of parsec around the energetic ($\dot{E} \geq 10^{34}$ erg/s) 
pulsars, and well beyond the boundaries of the pulsar wind nebulae observed in the radio and X-ray bands, see e.g.~the case of Geminga \citep{Posselt:2016lot,Pellizzoni_2011}. 
This extended emission can be explained as originating from energetic electrons and positrons (collectively referred to as electrons hereafter) accelerated by the pulsar and producing inverse Compton scattering (ICS) emission off the ambient radiation fields after escaping the relic pulsar wind nebula, see e.g.~\cite{HAWC:2017kbo,DiMauro:2019yvh,DiMauro:2019hwn,Tang:2018wyr,Martin:2022hrx}. However, an  
exhaustive theoretical understanding of the process, and in particular of the inferred diffusion properties in the vicinity of the pulsar, is still missing \citep{Sudoh:2019lav,Giacinti:2019nbu,Martin:2022hrx}. 
Notably, the spectral and morphological properties of these halos suggest that the diffusion of electrons is inhibited with respect to what is found to describe cosmic ray propagation in our Galaxy. 
Despite a number of mechanisms that have been proposed to obtain the low particle diffusion environment in pulsar halos \citep{Evoli:2018aza,Mukhopadhyay:2021dyh,Schroer:2022gau,Lopez-Coto:2022igd,Liu:2022hqf,Fang:2022fof,Recchia:2021kty}, at present none of them are able to consistently explain the properties of the halo around the Geminga pulsar. 
Understanding the particle emission and the escape mechanism of electrons and positrons from Geminga and other Galactic pulsars  is fundamentally important to constrain their contribution to the excess of positrons observed in the local Galactic flux \citep{Manconi:2020ipm,Martin:2022hrx,Lopez-Coto:2022igd}. 

If the gamma-ray halos originate from a distribution of highly energetic electrons, synchrotron losses of the same electrons in the ambient magnetic fields are expected to produce a diffuse emission with a similar spatial extension. 
Such putative synchrotron halos can provide independent measurements of the interstellar magnetic field strength in which electrons propagate, one of the crucial ingredients for theoretical modelling of particle diffusion in halos~\citep{Liu:2022hqf}. Moreover, multiwavelength study of the synchrotron and gamma-ray halo can place stringent constraints on the properties of the energetic electron populations including the cutoff energy of their injection spectrum. 

Given the large size of the gamma-ray emission ($\sim$ few degrees) and the expected low interstellar magnetic field (1--3 $\mu$G), the synchrotron counterpart of the Geminga pulsar halo is predicted to be largely extended and extremely faint. Since the leptonic nature of the halo predicts that the electron distribution is more concentrated around the pulsar wind nebula at higher energies, probing the synchrotron emission from the most energetic electrons in the X-ray band may provide higher detection  probabilities. Nevertheless, the attempts to search for Geminga's synchrotron halo in the X-ray band (\cite{Liu:2019sfl} using \xmm{} and Chandra, \cite{Khokhriakova:2023rqh} using SRG/\textit{eRosita}) yielded non-detection with the estimated magnetic field of sub-$\mu$G. These attempts bear limitations such as a narrow energy and spatial coverage (\cite{Liu:2019sfl}, see \S\ref{sec:xmm_data}) and short (400 s) exposure \citep{Khokhriakova:2023rqh}. 

In this paper, we present the most comprehensive search to date for the X-ray counterpart of the Geminga pulsar halo over a broad energy range (0.5 -- 79~keV) using \xmm{} and \nustar{}. We adopt novel analysis techniques to (1) accurately measure the background over the entire field of view (FoV), and (2) utilise both focused and stray-light photons to overcome the limited FoV of the X-ray instruments ($\leq 0.5^{\circ}$) and extend the observable region out to $\theta \sim 4^{\circ}$.
To investigate the X-ray halo profile robustly, we subtract the contribution of the Geminga pulsar, its wind nebula and diffuse X-ray background components. 
%

%
The paper is organised as follows. 
In Sect. \ref{sec:theory} we describe our theoretical framework for the prediction of 
synchrotron diffuse halos around pulsars.  In Sect. \ref{sec:X-ray_data} we present the analysis of the {\xmm} and {\nustar} archival data. 
In sect. \ref{sec:results} we derive X-ray halo flux limits in different energy bands, and discuss the consequences on model parameters, in particular the magnetic field. We also correlate the synchrotron emission with the ICS one and provide an interpretation of the gamma-ray data from HAWC, H.E.S.S. and \fermi{}-LAT.  
We draw our conclusions in Sect. \ref{sec:conclusions}.

\section{Synchrotron emission in Geminga's pulsar halo}
\label{sec:theory}
Electrons with energies 
$E$ emit synchrotron radiation in the ambient magnetic field $B$ at energies of about 
$E_{\rm sync}\sim 40\,{\rm keV} \frac{B}{\mu{\rm G}} (\frac{E}{\rm PeV})^2$, 
meaning that sub-PeV electrons are expected to emit photons at keV energies if $B$ is a few $\mu$G. 
This implies that a synchrotron counterpart of the Geminga's pulsar ICS halo detected at photon energies of tens of TeV is expected to emit diffuse X-rays at keV energies and at few-degree angular scales.
 Understanding the expected spectral and spatial properties of Geminga's pulsar halo at keV energies is crucial to defining the search in X-ray datasets reported in the next sections. 
In this section we review the modelling of the electron propagation and non-thermal emission in Geminga's pulsar halo (section \ref{subsec:model}) and we illustrate the spectral and spatial properties of the expected emission from keV to TeV energies (section \ref{subsec:showcase}).

\subsection{Electron propagation and multiwavelength emission modelling}
\label{subsec:model}
We model the  multiwavelength, non-thermal emission of electrons around Geminga following the procedures described in~\cite{DiMauro:2019yvh,DiMauro:2019hwn,DiMauro:2020jbz}, specifically devoted to the ICS emission from energetic electrons. 
The energy spectrum of accelerated electrons  is considered to be in the form of a power-law model with an exponential cutoff $Q(E)=Q_0 E^{-\gamma} \exp(-E/E_c)$. The normalisation of the energy spectrum $Q_0$ is determined by relating the total energy emitted in electrons  by Geminga to the total spin-down energy, assuming  a given efficiency $\eta$ of conversion of the spin-down luminosity into electrons and positrons pairs, see Eqs.9-12 in \cite{DiMauro:2019yvh}. 
Other relevant parameters for the Geminga's pulsar halo emission are the age, distance, and spin-down properties of the pulsar. 
We fix a distance of 250~pc, age of 342 kyr, current spin-down power of $3.2 \times 10^{34}$~erg s$^{-1}$, a magnetic dipole braking index of $n=3$ and a typical decay time of $\tau_0=12$~kyr, 
being the spin--down time evolution 
$\dot{E}(t) = \dot{E}_0 ( 1+ \frac{t}{\tau_0})^{-\frac{n+1}{n-1}}$.

We then solve the transport equation for electrons taking into account diffusion and energy losses by synchrotron emission (similarly for ICS), in order to compute the electron density at different positions around the pulsar. 
The electron density at distance  $\mathbf{r}$ from the pulsar is then integrated along the line of sight to obtain the observed flux from synchrotron emissions. 

Following the recent works, in order to explain the morphology and energy dependence of the gamma-rays observed by HAWC, H.E.S.S. and \fermi{}-LAT, 
we assume that diffusion around Geminga is inhibited with respect to the mean value found by fitting Galactic cosmic ray nuclei, see e.g. \cite{HAWC:2017kbo,DiMauro:2019yvh,HESS:2023sbf}. 
The diffusion coefficient is defined as $D(E)=D_0 (E/1 \textrm{GeV})^\delta$, where $D_0$ is the value at a reference energy of 1 GeV and we fix $\delta=0.33$ as previously done in \cite{DiMauro:2019hwn,Recchia:2021kty}.
To model the ICS emission at GeV energies, which extends a few tens of degrees (corresponding to a few tens of parsec) around Geminga, we use a two-zone diffusion model, in which  the inhibited diffusion is confined to a region of radius $r_b$ \citep{Tang:2018wyr,DiMauro:2019hwn} (see also \cite{Osipov:2020lty}). 
Outside of this region, electrons are propagated assuming the typical diffusion properties as constrained by cosmic ray nuclei within a similar semi-analytical propagation setup \citep{Kappl:2015bqa}. 
In addition to diffusion, radiative cooling of electrons by means of ICS in the local interstellar radiation field (ISRF) and synchrotron losses are considered. We implement the full Klein-Nishina cross section for the ICS losses as described in \cite{DiMauro:2019yvh}.  

The synchrotron photon flux emitted from Geminga pulsar halo  at an energy $E_{\gamma}$ and for a solid angle $\Delta \Omega$ is computed as:
\citep{1970RvMP...42..237B,Cirelli:2010xx}:
\begin{equation} \label{eq:phflux}
 \phi^{\rm} (E_{\gamma}, \Delta \Omega)= \frac{1}{4\pi} \int_{m_e c^2}^{\infty} dE \mathcal{M}(E, \Delta \Omega) \mathcal{P}^{\rm sync}(E, E_{\gamma})\,.
\end{equation}
The term $\mathcal{M}(E,\Delta \Omega)$ is the spectrum of electrons of energy $E$  in the solid angle $\Delta \Omega$:
\begin{equation}
\mathcal{M}(E,\Delta \Omega)  = \int_{\Delta \Omega} d\Omega \int_0^{\infty} d s \, \mathcal{N}_e (E,s).
 \label{eq:M}
\end{equation}
being $\mathcal{N}_e (E,s)$ the energy spectrum of electrons and positrons of energy $E$,  $s$ the line of sight, and  $\mathcal{P}^{\rm sync}(E,E_{\gamma})$ the power of photons emitted by a single electron for synchrotron emissions. We define the solid angle $\Delta \Omega$ by using the angular separation between the line of sight $s$ and the angular direction of the source $\theta$ in the sky. 
The parameter $\theta$ is highly relevant when comparing the model prediction to data, as it determines the amount of synchrotron (or inverse Compton) emission contained within the FoV of observation by different instruments. 

The synchrotron power  $\mathcal{P}^{\rm Sync}(E,E_{\gamma})$ is defined as \citep{2010PhRvD..82d3002A}:
\begin{equation}\label{eq:syncspectrum}
 \mathcal{P}^{\rm Sync}= \frac{dN_{\rm Sync}}{dE_{\gamma} dt}\,. 
\end{equation}
We connect the quantity defined in Eq.~\ref{eq:syncspectrum} to the energy emitted by one electron or positron per unit frequency and unit time,  $\frac{dE_{\rm sync}}{d\nu dt}$, as:
\begin{equation}
 \frac{dN_{\rm Sync}}{dE_{\gamma} dt}= \frac{1}{h E_{\gamma}} \frac{dE_{\rm sync}}{d\nu dt}
\end{equation}
since $N_{\rm sync} E_{\gamma}= E_{\rm Sync}$.
In a random magnetic field, the emissivity function is obtained by averaging out the standard synchrotron formula (see \cite{1970RvMP...42..237B}) over the directions of the magnetic field. 
The emitted energy per unit frequency in the case of electrons with  arbitrary pitch angle is given by  \citep{2010PhRvD..82d3002A}:
\begin{equation}
 \frac{dE_{\rm sync}}{d\nu dt} = \frac{\sqrt{3} e^3 B}{m_e c^2} G(x)
\end{equation}
being $e$ and $  m_e$ the electron charge and mass, respectively, $B$ the magnetic field and $c$ the speed of light. 
We use the function $G(x)$  as defined in \cite{2010PhRvD..82d3002A} (Eq.~D7), which is a precise analytical approximation for the dimensionless synchrotron integral, where $x=\nu/\nu_c$, $\nu=E_{\gamma}/h$ and
\begin{equation}\label{eq:nnuc}
 \nu_c= \nu_c(E)= \frac{3e B E^2}{4\pi m_e^3 c^5}\,. 
\end{equation}

We note that our computations assume that the synchrotron emission is produced in a uniform, random ambient magnetic field of value $B$ extending at least to the scale of the observed gamma-ray ICS emission, as done in many other recent works.
This is expected to be an approximation, as the magnetic field around the Geminga pulsar wind nebula could have a more complicated structure, also depending on the diffusion properties \citep{Liu:2022hqf}. 
In turn, the investigation of pulsar synchrotron halos using X-ray data with better angular resolution could potentially reveal morphological signatures informing the underlying magnetic field structure. 
In the absence of a significant detection of a synchrotron halo emission, the obtained upper limits are interpreted assuming the uniform B-field within the halo which reproduces the observed gamma-ray data, leaving the investigation of more complicated magnetic field structures to future work.

The multwavelength ICS SED and morphology models are calculated as described in \cite{DiMauro:2019yvh,DiMauro:2019hwn}, to which we refer for any further details. In particular, we model the ISRF following the local model of \cite{Vernetto:2016alq}. 

We finally note that the proper motion of Geminga (transverse speed $v_T\approx 211$(d/250pc) km s$^{-1}$ \citep{Hobbs:2005yx}) has been found to imprint an asymmetry in the pulsar halo \citep{Tang:2018wyr,DiMauro:2019hwn,Johannesson:2019jlk}, detected within the GeV halo emission with \fermi{}-LAT data \citep{DiMauro:2019yvh}. The asymmetry produced at TeV energies is instead predicted and observed to be negligible, due to the energy loss timescale. Given our focus on $\sim$ keV photons, which correspond to the same population of TeV-emitting electrons, we neglect the effect of the proper motion on the synchrotron pulsar halo. We explicitly verified that no observable distortion is expected within \xmm{} and \nustar{} energy ranges. 
For a general investigation of the effects of the pulsar's proper motion on the properties of pulsar halos, see ~\cite{DiMauro:2019yvh,Zhang:2020vga}.

\subsection{Spectral and spatial properties}\label{subsec:showcase}

In this section we illustrate the spectral and spatial properties of the Geminga halo emission, focusing on the expected characteristics in the X-ray band. 
Unless otherwise stated,  we show results using the following benchmark parameters: $\gamma=1.9$, $E_c=10^3$~TeV, $\eta=0.16$, one zone diffusion with $D_0=10^{26}$~cm$^2$s$^{-1}$, and a magnetic field of $B=3 \mu$G, which are very similar to those used in many recent works to interpret Geminga's halo at GeV-TeV energies, see e.g. \cite{DiMauro:2019yvh,HESS:2023sbf}. These parameters will be further checked in section \ref{sec:interpretation} for their consistency with ICS emission data at different angular scales. 
We refer also to \cite{Li:2021nrm} for complementary investigations on the properties of synchrotron pulsar halos. 

\begin{figure*}
    \centering
    \includegraphics[width=0.49\linewidth]{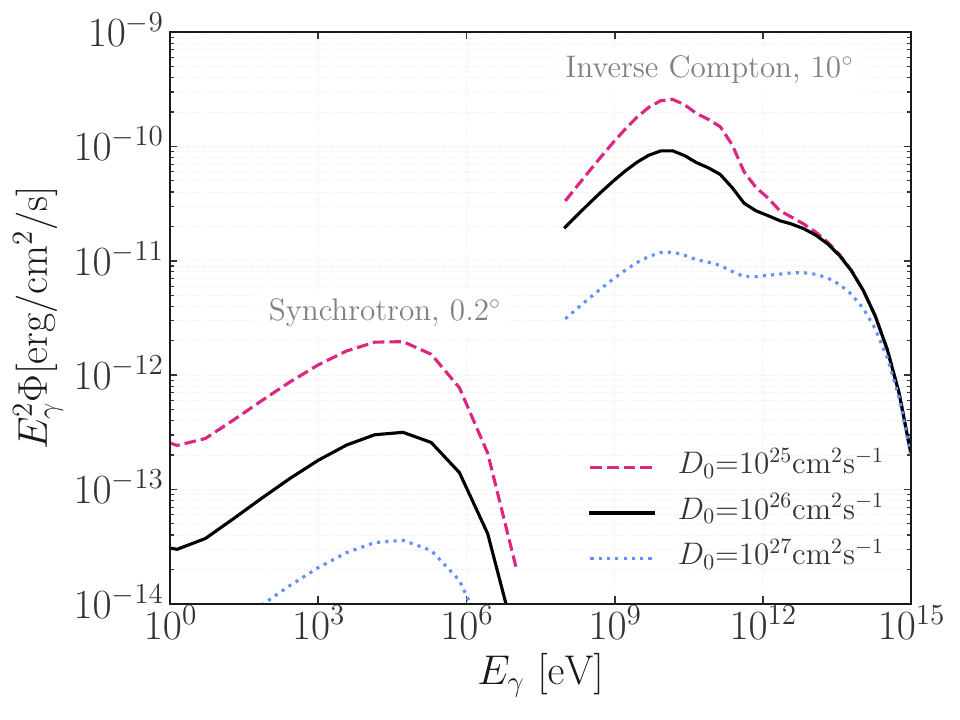}
\includegraphics[width=0.49\linewidth]{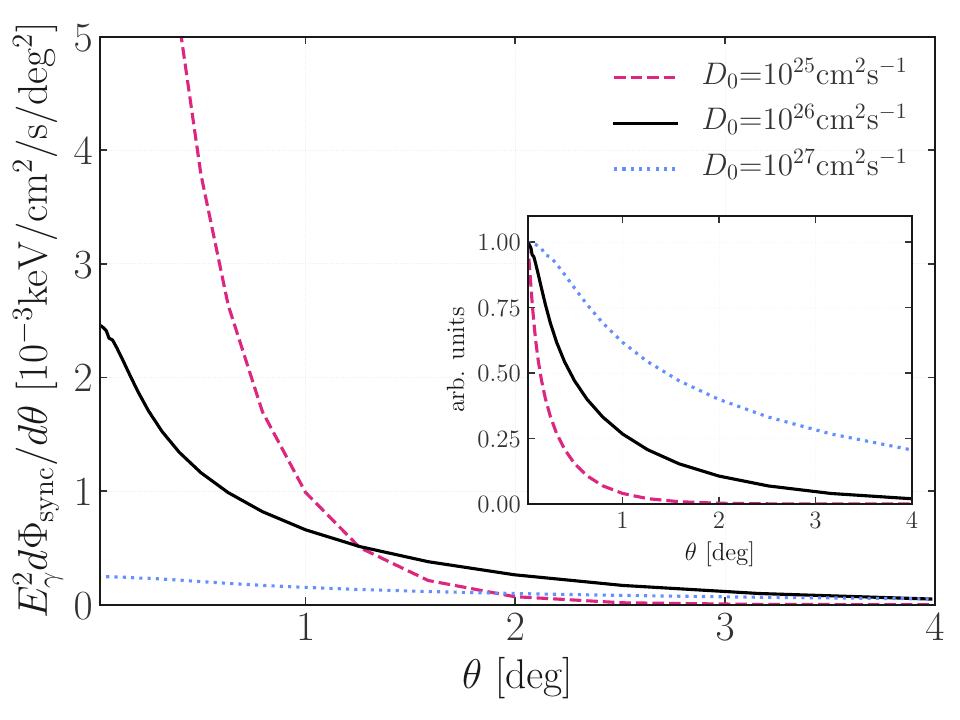}
    \caption{\textit{Effect of the diffusion coefficient on the Geminga's halo properties}. Left: spectral energy distribution of the Geminga pulsar halo for the synchrotron and ICS emissions, as integrated within an angular radius of $0.2^{\circ}$ and $10^{\circ}$, respectively, for different values of the diffusion coefficient at $1$~GeV ($D_0$).  
    Right: surface brightness of the Geminga synchrotron halo integrated over the energy range $E_{\gamma}$=[10--40]~keV for different $D_0$. The inset shows the flux profile normalised at $\theta=0$, in arbitrary units. 
    }
    \label{fig:D0effect}
\end{figure*}

The diffusion properties for electrons accelerated by the pulsar and injected within the surrounding interstellar medium are crucial to determine the halo angular extension and the normalisation of the observed flux within a given observation angle, both at gamma-ray and X-ray energies. 
We refer to \cite{DiMauro:2019hwn} for an extensive investigation of the expected angular extension of the ICS emission as a function of the distance and age of the pulsar for suppressed diffusion, and to \cite{Liu:2022hqf} for a review of the current challenges in understanding the nature of the propagation properties in pulsar halos. 
As found in the first observation of the Geminga halo at TeV \citep{HAWC:2017kbo}, in order to explain the surface brightness of the observed few-degree extended emission, a diffusion coefficient $D_0\sim10^{26}$~cm$^2$s$^{-1}$, lower by a factor of about 500 with respect to the mean value for the rest of the Galaxy, is required. 
For higher values of the diffusion coefficient, the Geminga's halo at TeV would be widespread at much larger angular scales, while values towards $D_0\sim10^{25}$~cm$^2$s$^{-1}$ would concentrate even more the emission within the inner few degrees.
We study the dependence of the flux and of the surface brightness of Geminga in Fig.~\ref{fig:D0effect}, keeping in mind that the same electrons producing ICS TeV gamma-rays emit keV synchrotron photons. 
In the right panel, we plot the surface brightness as integrated in the energy bin 10--40~keV for different values of $D_0$. 
Similarly to what was found for the ICS halo, very low values of $D_0$ give an X-ray halo highly concentrated towards the inner $\sim1^{\circ}$  around the Geminga pulsar, while higher values of $D_0\sim10^{27}$ cm$^2$s$^{-1}$ distribute the emission over larger angular scales, making a significant detection more difficult. 
Another representation of this effect is shown  in the inset figure, showing the flux profile normalised at $\theta=0$ in arbitrary units. 
At fixed conversion efficiency, and integrating within the same angular scales $\theta$ of each pulsar halo emission, this translates into an overall normalisation effect to the spectral energy distribution (SED), shown in the left panel of  Fig.~\ref{fig:D0effect}. 
We illustrate the ICS emission integrated within the inner $ 10 ^{\circ}$, and the synchrotron emission integrated within the inner $0.2^{\circ}$, the latter corresponding roughly to the \nustar{} FoV, see next Section. 
For the chosen benchmark parameters, the synchrotron emission is peaked at tens of keV, and quickly decreases at higher energies, given the exponential cutoff in the electron source spectrum. 
A change in $D_0$ by an order of magnitude shifts the overall normalisation of the synchrotron emission by a similar amount. 
We understand this behaviour as linked to the integration angle. In the very small region ($0.2^{\circ}$) considered for the synchrotron emission, source electrons are diffusion-dominated. On the wide ICS angular size, instead, the propagation of electrons is also controlled by energy losses. 
At GeV-TeV energies, the gamma-ray surface brightness and SED data for Geminga imply that  $D_0$, $\eta$ and the spectral parameters are highly correlated, see e.g. the discussion in \cite{DiMauro:2019yvh}.

\begin{figure*}
    \centering
    \includegraphics[width=0.49\linewidth]{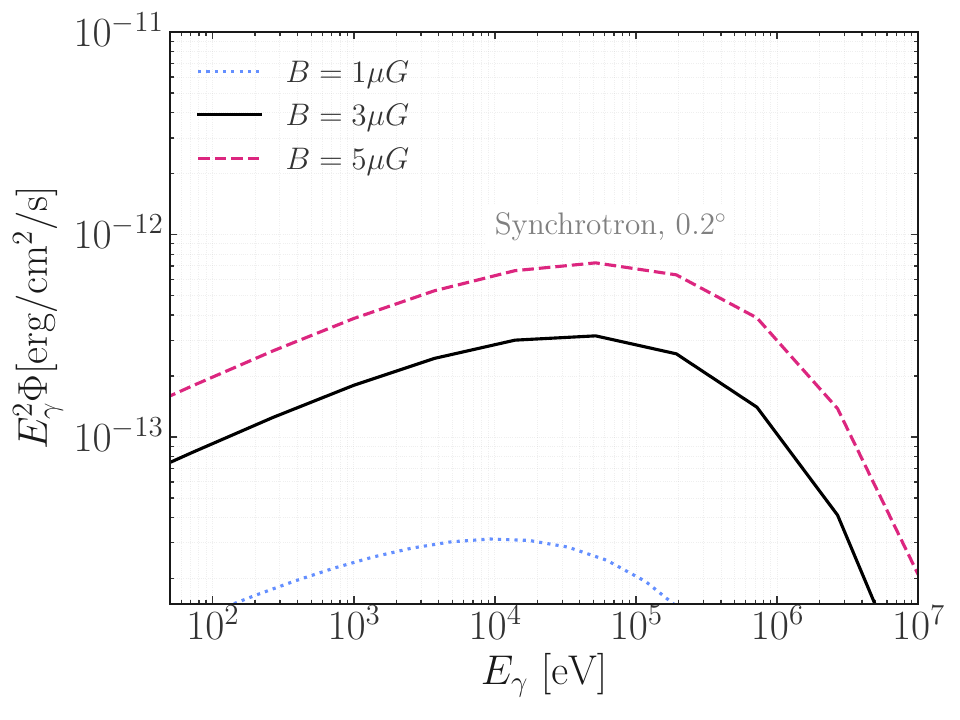}
\includegraphics[width=0.49\linewidth]{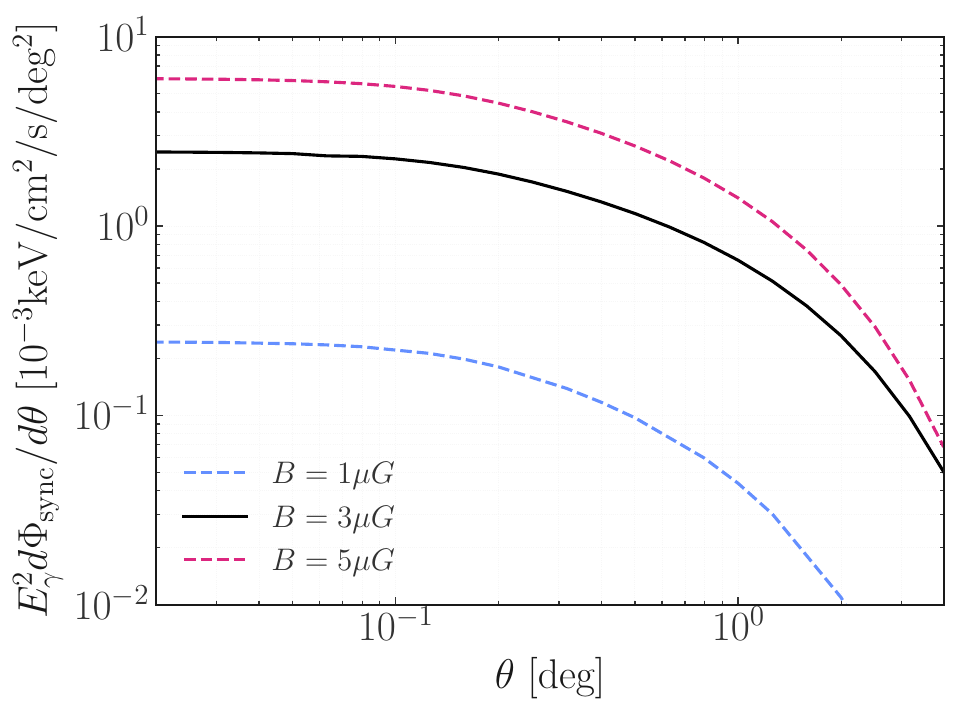}
    \caption{\textit{Effect of the magnetic field value on the properties of Geminga's synchrotron halo}. Left: SEDs of the Geminga pulsar halo for the synchrotron emissions integrated over an angular radius of $0.2^{\circ}$ for different values of the magnetic field $B$. 
    Right: surface brightness of the Geminga synchrotron halo integrated over the energy range $E_{\gamma}$=[10--40]~keV for different $B$. Note the double log scale. }
    \label{fig:Beffect}
\end{figure*}

The effect of the magnetic field is displayed in  Fig.~\ref{fig:Beffect} for the 
synchrotron SED and the surface brightness. The SED peak position scales with B, while the normalisation of the SED at its peak scales roughly with $B^2$, which follows the total energy loss rate. A similar effect is observable for the surface brightness (right panel), which implies an integration over the energies from 10 to 40 keV. 

\begin{figure*}
    \centering
        \includegraphics[width=0.49\linewidth]{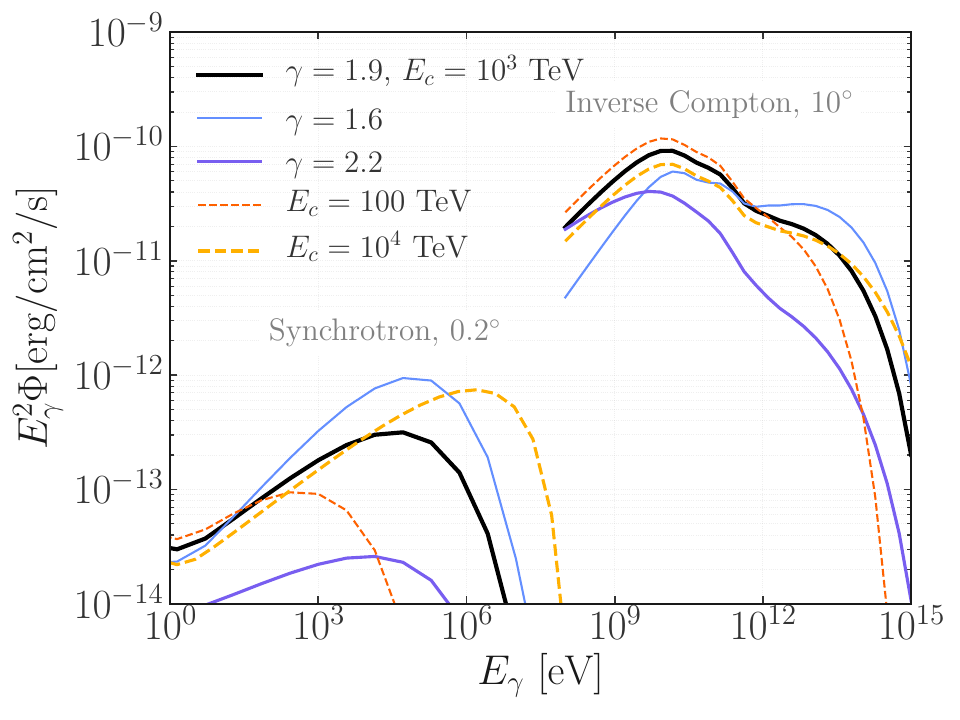}
\includegraphics[width=0.49\linewidth]{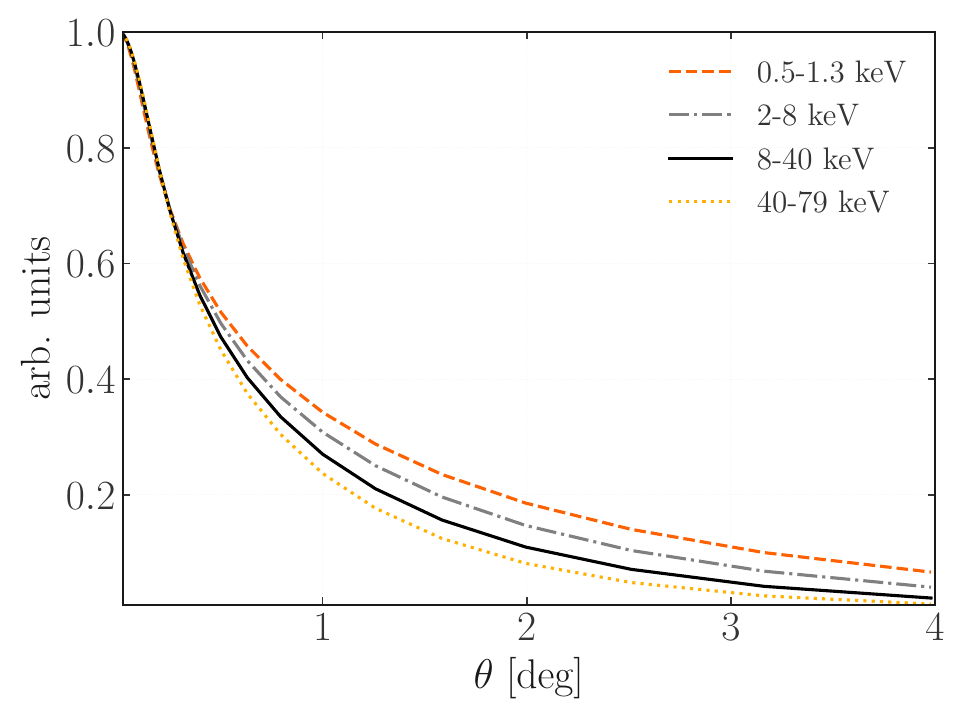}
    \caption{\textit{Left panel:} SEDs of the Geminga pulsar halo emission for different values of the spectral parameters $\gamma$ and $E_c$ of the electron's source term $Q(E)\propto E^{-\gamma} \exp{(-E/E_c)}$. When not specified, curves are obtained for the benchmark value of $\gamma=1.9$ and $E_c=10^3$~TeV.  Angular integration as in Fig.~\ref{fig:D0effect}. \textit{Right panel:} Surface brightness of the Geminga's pulsar halo obtained by integrating in different energy intervals, corresponding to the {\xmm} (0.5-8~keV) data analysis, and to the {\nustar} ones. All curves are rescaled to their value at $\theta=0$.  }
    \label{fig:spectra_sb}
\end{figure*}

The left panel of Fig.~\ref{fig:spectra_sb} illustrates the effect of changing the spectral properties of the electrons undergoing ICS and synchrotron emission, by means of the spectral index $\gamma$ and the cutoff energy $E_c$ of the $Q(E)$ defined above, again at fixed $\eta$. 
We observe that a soft spectral index of $\gamma = 2.2$ largely suppresses both emissions in the considered energy range.  
As for the cutoff, a value of $E_c=100$~TeV would suppress the synchrotron emission at tens of keV and at tens of TeV, in tension with the observation by HAWC and LHAASO \cite{HAWC:2017kbo}, while a higher value of $10^4$~TeV would shift the peak of the synchrotron emission to hundreds of keV. 
Finally, the right panel of Fig.~\ref{fig:spectra_sb} shows the synchrotron surface brightness profile as integrated over different energy bands, corresponding to the \xmm{} and \nustar{} observations discussed in the next section. 
For the benchmark configuration, the surface brightness of 8--40~keV photons is expected to decrease by about a  factor of two in the inner $\theta = 0.25^{\circ}$. 
By normalising all the curves to their value at $\theta=0$, we observe that the halo profile gets more concentrated around the pulsar at higher energies. 
This is expected, similarly to the energy dependence seen from the ICS gamma rays: higher energy photons are produced by higher energy electrons, which are mostly  distributed closer to the pulsar, before diffusing away and losing energy.

\section{X-ray analysis}\label{sec:X-ray_data}
We analysed archival \xmm{} (0.5 -- 8 keV) and \nustar{} (8 -- 79 keV) data to cover a broadband spectrum of the Geminga halo. Since the very faint emission of the halo is expected to span the entire FoV of both instruments, careful background estimation is critical in this analysis. The background of each instrument is dominated by distinct components due to their specific energy range and instrumental design, and hence, different approaches are required for estimation. We first present our analysis of the \xmm{} data in \S \ref{sec:xmm_data} that represent the focused X-rays in the soft X-ray band in the inner $\theta = 0.25^{\circ}$ of the Geminga halo thanks to its highly efficient stray light rejection and wide FoV. We describe our novel technique to estimate various particle backgrounds, the most significant background component for \xmm{}. In \S \ref{sec:nustar_data}, we present the \nustar{} analysis of both focused ($\theta = 0.1^{\circ}$) and stray light ($\theta = 1^{\circ} - 4^{\circ}$) photons from the Geminga halo in the hard X-ray band. In general, the \nustar{} observations are heavily influenced by stray light photons from the cosmic X-ray background emission, nearby bright point sources and ambient diffuse emission. While no bright X-ray point sources are present near Geminga, the emission from the Geminga halo itself may have a higher contribution to the data in the form of stray light photons than focused photons. We describe our novel analysis method to utilise both focused and stray light photons from a source with an extension much greater than the \nustar{} FoV. We adopt these new techniques to place stringent upper limits on the faint synchrotron halo of Geminga to be used to constrain our model parameters (magnetic field and diffusion coefficient) in \S \ref{sec:interpretation}.

\begin{table*}\caption{\nustar{} and \xmm{} observations of the Geminga pulsar ROI used for our analysis.}\label{tab:gemingaobs}
\centering 
\begin{tabular}{lccr} 
\hline \hline
Instrument & {Date}   & {ObsId} & {Exposure [s]} \\
\hline 
\xmm{} & 2002-04-04 & 0111170101 & 149,834 \\
& 2004-03-13 & 0201350101 & 33,495 \\
& 2006-03-17 & 0311591001 & 22,185 \\
& 2006-10-02 & 0400260201 & 37,422 \\
& 2007-03-11 & 0400260301 & 39,627 \\
& 2007-09-18 & 0501270201 & 45,134 \\
& 2008-03-08 & 0501270301 & 24,411 \\
& 2008-10-03 & 0550410201 & 36,588 \\
& 2009-03-10 & 0550410301 & 21,026 \\
\hline 
{ \xmm{} total Exposure Time [s]} & 409,721  &&\\
\hline 
\nustar{} & 2012-09-25 & 30001029018 & 26,488 \\
& 2012-09-27 & 30001029028 & 23,331 \\
& 2012-09-26 & 30001029022 & 20,924 \\
& 2012-09-21 & 30001029012 & 13,759 \\
& 2012-09-20 & 30001029006 & 9,428 \\ 
\hline 
{\nustar{} total Exposure Time [s]} & 93,930  & &   \\
\hline \hline
\end{tabular}
\end{table*}

\subsection{\xmm{} data analysis}\label{sec:xmm_data}
\xmm{}, the X-ray Multi-Mirror Mission, is an X-ray space telescope placed in orbit by the European Space Agency. Its primary instruments, the three European Photon Imaging Cameras (EPIC), include two MOS-CCD cameras (MOS1 and MOS2) and a single pn-CCD camera (pn). Together, they offer sensitive imaging over a large FoV ($0.5^{\circ}$) with moderate angular resolution (FWHM $6\arcsec$) in 0.15 to 15 keV.

A total of nine archival \xmm{} observations pointed at the Geminga pulsar from 2002 to 2009 were analysed.
Data reduction was done using the Standard Analysis Software (SAS v.21.0.0) and the most up-to-date calibration files.
The EPIC-MOS data taken in Full-Frame mode was used in searching for emissions of the Geminga halo.
To remove the background flares, the mean and the variance of a period of low-rate data are calculated, and the data in the observation with a rate above the mean by more than 2.5 times of the standard deviation ($2.5\sigma$) were rejected.
After filtering out background flares,
we obtained a total exposure time of 409.7 ks.

For spectral analysis, source counts were extracted from the entire FoV of the EPIC-MOS camera, excluding the circular region with a radius of $0.015^{\circ}$ centred at the Geminga pulsar. 

The large FoV of \xmm{} offers not only a broad view of the inner region of the Geminga halo but also challenges in estimating the background over such a large region. The background of \xmm{} consists of four main components, namely quiescent particle background (QPB), soft-proton flare (SPF), diffuse X-ray background (DXB), and solar wind charge exchange (SWCX). Extensive studies on the background estimation have been incorporated into the \xmm{} Extended Source Analysis Software\footnote{\url{https://heasarc.gsfc.nasa.gov/docs/xmm/xmmhp_xmmesas.html}} (ESAS) for users. The ESAS is equipped with convenient tools to constrain the QPB component relatively well using the data outside of the FoV (corner data), and filter the time intervals with highly increased count rate due to SPFs. However, estimating the residual SP component after filtering and the DXB component rely on a spectral fitting of numerous line emissions as well as thermal and non-thermal continuum to the data. This is particularly difficult in cases where the extended source has (1) a very low count rate, and (2) supposedly a spectrum that resembles the background spectrum. Our analysis of the Geminga halo meets both conditions as the power-law emission of the halo is expected to be very faint. Moreover, estimating the uncertainties related to the distinct background components with distinct origins becomes increasingly important in case of a non-detection of the source to place stringent upper limits on the faint halo emission. We describe the novel data-driven techniques developed for our analysis to overcome the limitations of the existing techniques and provide accurate measurement of the flux and uncertainty of each background component. Our analysis covers the energy range of $[0.5,8]$ keV above which the instrument sensitivity drops significantly. The middle energy range $[1.3,2]$ keV was excluded because of the strong background line emissions.

\paragraph{Quiescent particle background.}
This background component is associated with high-energy particles interacting with the instrument structure.
The spectral and spatial distribution of QPB events was modelled using the Filter-Wheel-Closed (FWC) data taken during the CLOSED filter observations, and the corner data collected by the CCD pixels outside of the FoV during the science observations.
The FWC data was normalised to the science data using the count ratio (cts px$^{-1}$ s$^{-1}$) of the corner data between the observations. The normalized FWC spectral and spatial distributions are then fixed in the following fitting procedure that determines the SPF background.
\paragraph{Soft-proton flare.}
This background component arises from low-energy protons interacting with the detectors.
The distribution of SPF events was estimated using the event patterns.
It was empirically found that the pattern distribution of SPF events is distinct from the pattern distributions of X-ray events and QPB events.
The template of SPF pattern distribution is constructed using the data taken during the strong SP flares, while the template of the X-ray pattern distribution can be constructed using the data containing a bright X-ray source. The QPB template is determined using the FWC and corner data and held fixed. The templates of X-ray and SPF are then fitted to the Geminga data pattern distribution to determine the normalisation for the X-ray and SPF components.
\paragraph{Diffuse X-ray background.}
The X-ray component contains the X-ray events from the source as well as X-rays from the Galactic and extragalactic diffuse backgrounds (CXB and GRXE, see \S \ref{sec:nustar_data}). 
The DXB is measured using 13 observations within an angular distance range of $[\theta_{\mathrm{min}},\theta_{\mathrm{max}}]$ degrees from the location of Geminga pulsar, where $\theta_{\mathrm{min}}=1^{\circ}$ to reject strong contamination from the Geminga halo in the background samples, and $\theta_{\mathrm{max}}=15^{\circ}$ to ensure that the DXB in the selected background samples is representative of the Geminga observations. The observations contain no X-ray sources or a single point source, in which case the source was excised. After subtracting the QPB and SPF backgrounds, the DXB is measured by taking the average differential flux (keV$^{-1}$ cm$^{-2}$ s$^{-1}$ deg$^{-2}$) of the entire FoV (excluding point sources).
\paragraph{Solar Wind Charge Exchange.}
The Solar Wind Charge Exchange (SWCX) background originates from the collisions between the ions in the solar wind and the exospheric hydrogen near the Earth or the neutral interstellar medium.
The majority of SWCX is rejected by the count-rate-based filter, and residual of SWCX can be checked by the spectral analysis.
The SWCX emissions are most prominent from C VI, O VII, O VIII, Ne IX, and Mg XI lines.
These emission lines were not seen in the Geminga halo spectrum after the subtraction of QPB, SPF, and DXB components.
Therefore the SWCX is concluded to be negligible and was not investigated further.

The flux and uncertainty of each background component in the energy range of [0.5,8.0]~keV are summarised in Table \ref{tab:xmm_background}.
The QPB flux uncertainty originates from the statistical error of the corner data used to normalise the FWC data.
The SPF flux uncertainty comes from the statistical fluctuation of the data pattern distribution.
The DXB flux uncertainty is estimated as the standard deviation from the mean DXB flux of the 13 background samples.
We report in the appendix the detailed results on the DXB flux measurement of the individual observations as a function of angular distance to the Geminga pulsar.

\begin{figure*}
\centering
\includegraphics[width=0.49\linewidth]{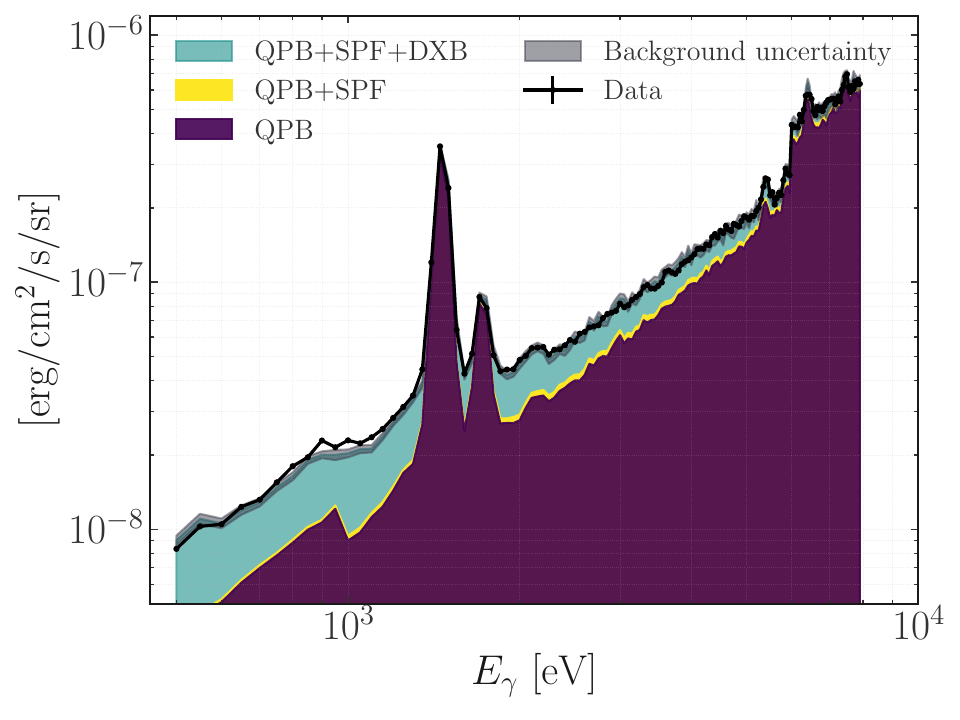}
\includegraphics[width=0.49\linewidth]{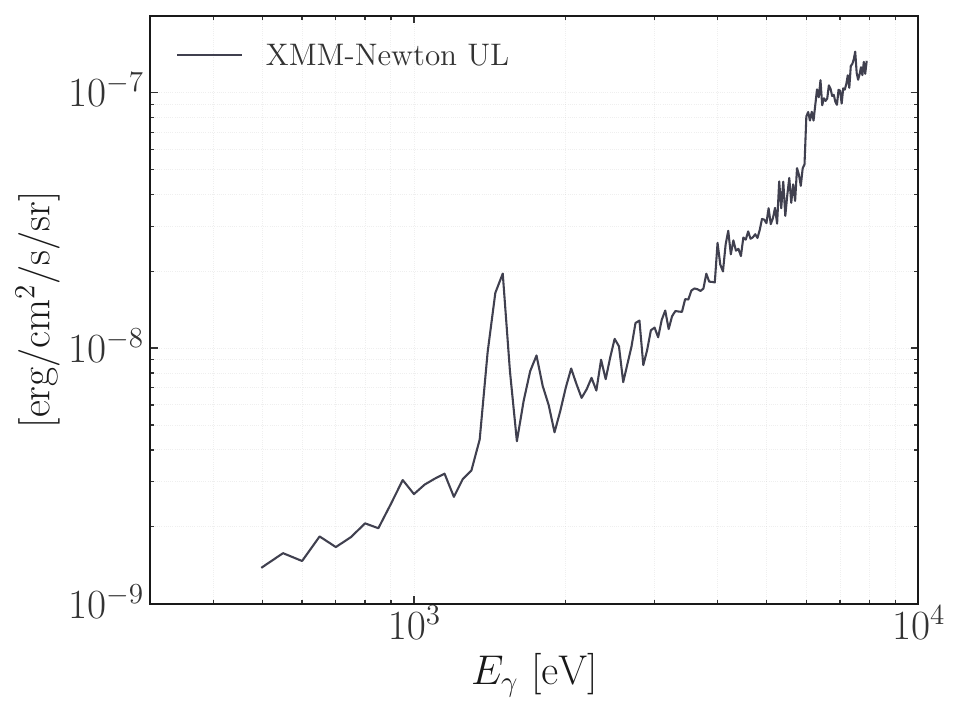}
\caption{ 
Left: The X-ray spectrum in $[0.5,8]$ keV range in \xmm{} data.
The data (black) is overlaid with the expected background components, including quiescent particle background (QPB, purple), soft-proton flare (SPF, yellow), and diffuse X-ray background (DXB, green).
The gray band shows the size of background uncertainty.
Right: The upper limit in the same energy range derived from the background uncertainty.
}
\label{fig:xmm_spectrum}
\end{figure*}

The predicted background components and the observed X-ray spectrum in $[0.5,8]$ keV are shown in Fig \ref{fig:xmm_spectrum} (Left).
The dominant backgrounds are QPB and DXB, while the SPF contribution is marginal after the initial filtering by the count rate ($2.5\sigma$ cut) that removed the majority of the SPF events.
The predicted background shows a good agreement with the observed data spectrum, indicating that the Geminga halo emission is insignificant compared with the background.
The 99\% upper limit for the Geminga halo emission is then calculated as three times the background uncertainty as shown in Fig.~\ref{fig:xmm_spectrum} (Right).

Our upper limits from the \xmm{} data provide strong constraints on the Geminga halo emission in the low-energy range ($[0.5,1.3]$ keV) but are less constraining in the high-energy range ($[2-8]$ keV) due to the strong QPB contribution. Thus, we calculated the upper limits in the two energy ranges separately for multiple annular regions around the Geminga pulsar. We multiply the differential upper limits by the solid angle of the FoV and the annular regions for the total flux upper limits.


We address different aspects of this work from the published analysis of the \xmm{} observation of Geminga by \cite{Liu:2019sfl}. 
First, a narrower energy range of 0.7 - 1.3~keV and a smaller region of interest of 600" radius were considered in \cite{Liu:2019sfl}, while this work covered an energy range of 0.5 - 8 keV and a region of interest of 900" radius.
Second, the null detection of the Geminga halo in \cite{Liu:2019sfl} is based on the observation of the flux as a function of angular distance from the pulsar, which shows no radial variation.
While the radial dependence of the flux can be used to test the Geminga halo emission profile variation within the \xmm{} FoV, it cannot rule out the scenario in which the halo emission is uniform over the FoV and extends beyond the FoV.
The background estimation in this work is derived from the control regions independent of the signal region (Geminga FoV data) and hence allows a measurement of the absolute flux of diffuse X-ray emission.
Third, the calculation of the upper limit in \cite{Liu:2019sfl} assumes only statistical fluctuation of the flux measurement. 
In this work, we consider a more realistic upper limit by including systematic uncertainties in the background estimation (see Table \ref{tab:xmm_background} and Fig.~\ref{fig:xmm}) in addition to the statistical uncertainty.

\begin{table*}\caption{Breakdown of \xmm{} background components averaged over [0.5,8.0] keV.}\label{tab:xmm_background}
\centering
\begin{tabular}{lccccc}
\hline \hline 
 & {Data} & {Sum background} & {DXB} &{QPB} &{SPF}\\
\hline 
Flux $(\mathrm{erg}/\mathrm{cm}^{2}/\mathrm{s}/\mathrm{sr})$ & $2.36\times 10^{-7}$ & $2.37\times 10^{-7}$ & $2.95\times 10^{-8}$ & $2.01\times 10^{-7}$ & $7.09\times 10^{-9}$ \\
Uncertainty $(\mathrm{erg}/\mathrm{cm}^{2}/\mathrm{s}/\mathrm{sr})$ & $5.83\times 10^{-9}$ & $1.44\times 10^{-8}$ & $1.04\times 10^{-8}$ & $9.95\times 10^{-9}$ & $6.72\times 10^{-10}$ \\
\hline 
\end{tabular}
\end{table*}

\subsection{NuSTAR data analysis}\label{sec:nustar_data}

The Nuclear Spectroscopic Telescope Array (\nustar{}) is a space-based X-ray telescope that consists of two co-aligned focal plane modules (FPMA and FPMB, FPMs hereafter) with nearly identical optics and detectors  \citep{Harrison2013}. The \nustar\ optics can focus incident X-rays in 3--79 keV with an angular resolution of 14\asec\ (FWHM), while the focal plane detectors can register X-ray photons up to $\sim160$ keV \citep{Krivonos2021,2022ApJ...941...35M}. \nustar{}'s broad hard X-ray coverage is uniquely suited for exploring synchrotron radiation from TeV -- PeV energy electrons in the interstellar magnetic field.

Incident X-ray photons are reflected twice through the optics and focused on the detector plane. The FoV for these focused, or 2-bounce, photons is $13'\times13'$.  
Due to the open geometry of the \nustar\ telescope, FPMs can also register unfocused X-ray photons from far off-axis angles of $\theta \sim 1\rm{-}4^{\circ}$ \citep{2017Madsen}. These stray-light, or 0-bounce, X-rays can hit the detectors directly from the side unless they are blocked by the aperture stops or optical benches \citep{2014Wik}. While most \nustar\ observations have been utilised for resolving X-ray sources with the sub-arcminute resolution optics, stray-light X-ray components have been used for investigating particular astrophysical sources such as the cosmic X-ray background \citep[CXB; ][]{Krivonos2021}, the Galactic Bulge \citep{Perez2019} and bright X-ray binaries \citep{Grefenstette2021,2022ApJ...926..187B}. The focused and stray-light X-ray components have distinct FoVs ($13'\times13'$ vs.  $\sim1\rm{-}4^\circ$) and undergo different convolution processes from the sky to the detector plane. The efficiency of focusing X-rays onto the detector plane is largely determined by the optics point spread function (PSF) and effective area. On the other hand, stray-light X-rays reach the detector plane directly in a collimator mode. 

For a synchrotron X-ray halo around Geminga extending over $\theta \sim4^\circ$, \nustar\ will register both focused and stray-light X-ray photons. As described in the previous section,
the predicted X-ray halo distribution peaks around the pulsar (on-axis) and spreads over $\theta \geq 4^\circ$, possibly matching the extension of the halo detected at TeV energies \citep{HAWC:2017kbo}. X-ray emission from the inner $13'\times13'$  will be focused on the detector plane as 2-bounce photons, while X-ray emission from the outer $\theta \sim 1-4^\circ$ will be observed as a stray light component; see further discussion below. 
The non-uniformity of the X-ray halo in sky coordinates complicates the problem even further. The resulting observation is a combination of X-rays from different parts of the halo that cannot be disentangled from each other for comparison with the model. Instead, we take a reverse approach where we convolve our model with energy-dependent templates for focused and stray-light photons to reproduce the observed combination of both photons.  The following sections describe our \nustar\ observations and analysis methodology.

\subsubsection{NuSTAR observations and background subtraction}\label{sec:nustar_obs}

\begin{figure*}[ht]
\centering
\includegraphics[width=0.8\textwidth]{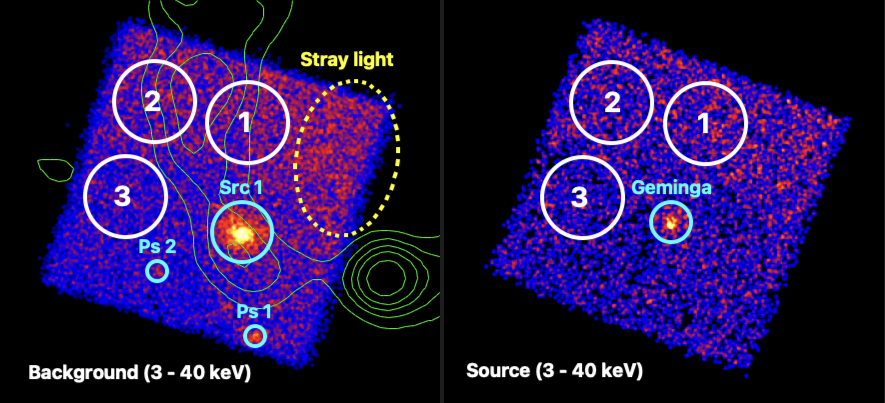}
\includegraphics[width=0.8\textwidth]{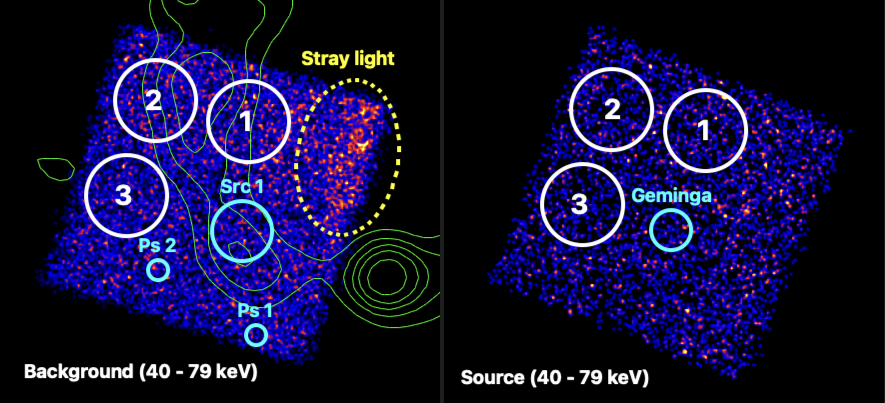}
\caption{Background and source counts map in two energy bands. The images were scaled for better visibility. Detected sources (Src 1, Ps 1 and 2, Geminga) are marked with cyan circles. Local stray light background only present in the background observation is marked with yellow dashed ellipse. VLA 74 MHz contour of IC 443 is overlaid in green (The VLA Low-Frequency Sky Survey Redux (VLSSr) Postage Stamp Server, \url{https://www.cv.nrao.edu/vlss/VLSSpostage.shtml}). Our regions of interest (region 1--3) are marked with white circles. }
\label{fig:nustar_images}
\end{figure*}

A series of 15 \nustar\ observations of the Geminga pulsar were carried out in 2012 \citep{Mori2014}. We analysed \nustar\ data from the five longest observations, as shown in Table~\ref{tab:gemingaobs}, with a total exposure of 94 ksec. In order to reduce the particle background, we processed the \nustar\ data using {\tt nupipeline} with {\tt SAAMODE=STRICT} and {\tt TENTACLE=yes}. Each observation with $\simgt10$ ks provides sufficient photon statistics, thus allowing us to investigate systematic differences between the observations. All five observations pointed at the Geminga pulsar on-axis with similar position angles (PA $\sim 157^\circ$). 
The Geminga pulsar and its compact PWN ($\theta\sim1'$) are visible as shown in Fig.~\ref{fig:nustar_images}.
No other bright X-ray sources are present in the FoV.

Below $\sim30$ keV, the background is dominated by stray-light X-rays whose pattern is non-uniform on the detector plane and solely dependent on the PA \citep{Mori2015}.  Above $\sim30$ keV, the internal detector background becomes increasingly prominent and is nearly uniform over the detector plane.
The sources of the stray-light background are (1) the CXB, (2) Galactic Ridge X-ray emission (GRXE), and (3) nearby bright X-ray sources. We confirmed that no X-ray sources that can cause significant stray light are present out to $\sim 4^{\circ}$.
While the CXB is isotropic, the GRXE should be negligible at the location of the Geminga pulsar ($l = 195.13^{\circ}$). Nevertheless, to account for the stray light from the local diffuse X-ray background, we utilised a nearby ($\sim6^\circ$ away from the Geminga pulsar) archival \nustar\ observation  (ObsID 30101058002, 102-ks exposure). The infrared brightness distribution (which traces the GRXE) between the pointing of this observation and Geminga shows no significant variation, indicating that diffuse X-ray background should be similar between the two regions. In addition, the observation was taken at a PA ($161^{\circ}$) nearly identical to those of the Geminga observations, allowing robust estimation of the stray-light background for Geminga.

The observation used for background estimation (``background observation") partially overlaps with the southeastern rim of the radio shell of the supernova remnant (SNR) IC 443 (green contour in Fig.~\ref{fig:nustar_images}).
The detection of an extended ($\sim 1'.5$ radius) source associated with the SNR (``Src 1") and two point sources (``Ps 1" and ``Ps 2") was reported by \citet{IC443}. No other significant X-ray emission is found in the FoV. The stray light pattern in the background observation is in good agreement with that of the Geminga observations (``source observations") up to 40 keV, demonstrating the PA-dependency of stray light background, and indicating that the stray light in the source and background observations are likely of the same origin (diffuse X-ray background). Above 40 keV, the background observation shows a distinct pattern of stray light of unknown origin in the northwest corner of the FoV that is not present in the source observations. We utilised the region away from this local stray light background and the detected sources for background estimation. We selected three circular regions of interest ($2'$ radius, region 1 -- 3 of white circles in Fig.~\ref{fig:nustar_images}) in the source and background observations such that each region is located at the same detector coordinates in all the observations. The background count rate for each region was extracted from the background observation, while the source count rate was extracted from the source observations. Note that the net count rate calculated from this background and source rate is the sum of the focused ($13' \times 13'$) and stray-light ($\theta \sim 1 - 4^{\circ}$ component of the putative Geminga X-ray halo. The net count rates in the regions are consistent with zero within the statistical uncertainties in 8--40 keV and 40--79 keV. We placed 99\% upper limits on the Geminga halo count rate for each region (Table \ref{tab:count_rates}). 

The largely negative count rates in 3--8 keV indicate background over-estimation. We attribute this over-estimation to two possible contributions: (1) the scattered X-ray emission from the Sun, which is most prominent below 5 keV and may differ between observations \citep{Krivonos2021}, and (2) very soft X-ray emission from IC 443 coincident with its radio shell. Since a more sensitive and reliable X-ray halo search below 10 keV can be conducted using soft X-ray telescopes such as \xmm{}, hereafter, we constrain the X-ray halo model parameters based on the count rate upper limits above 8 keV.

\begin{table*}
\caption{\nustar\ net count rates, 1$\sigma$ statistical errors, and 99\% upper limits on the net count rates in the three regions of interest marked in Fig.~\ref{fig:nustar_images}.}\label{tab:count_rates}
\begin{tabular}{lcccc}

\hline \hline 
{Energy band} & {} & {Region 1 [cts\,s$^{-1}$]} & {Region 2 [cts\,s$^{-1}$]} & {Region 3 [cts\,s$^{-1}$]}
\\
\hline 
3 -- 8 keV$^{\dagger}$ & Net count rate & $(-30.72 \pm 3.52) \times10^{-4}$ & $(-17.31 \pm 3.07) \times10^{-4}$& $(-16.02 \pm 2.84) \times10^{-4}$  \\ 
8 -- 40 keV & Net count rate & $(8.15 \pm 4.91) \times10^{-4}$ & $(7.43 \pm 4.77) \times10^{-4}$ & $(10.09 \pm 4.51) \times10^{-4}$\\
& 99\% UL & $2.29\times10^{-3}$ & $2.18\times10^{-3}$ & $2.36\times10^{-3}$ \\ 
40 -- 79 keV & Net count rate & $(5.35 \pm 3.99) \times10^{-4}$ & $(-0.65 \pm 3.94) \times 10^{-4}$ & $(5.66 \pm 3.72) \times10^{-4}$\\ 
& 99\% UL & $1.73\times10^{-3}$ & $1.18\times10^{-3}$ & $1.68\times10^{-3}$ \\ 
\hline 
\hline 
\multicolumn{5}{l}{$\dagger$ We do not report ULs in this energy band due to the two reasons elaborated at the end of \S \ref{sec:nustar_obs}.}
\end{tabular}
\end{table*}

\subsubsection{X-ray halo model convolution and flux upper limits} 
\begin{figure*}
    \centering
\includegraphics[width=0.43\linewidth]{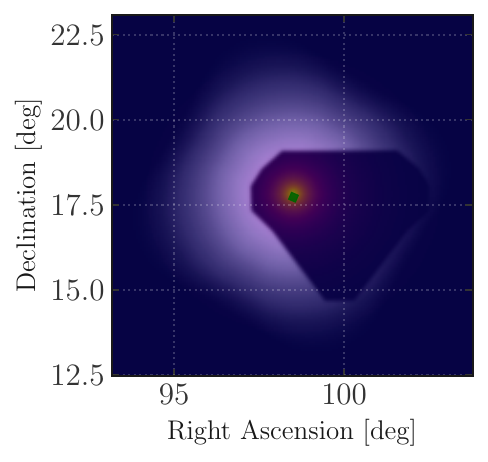}
\includegraphics[width=0.555\linewidth]{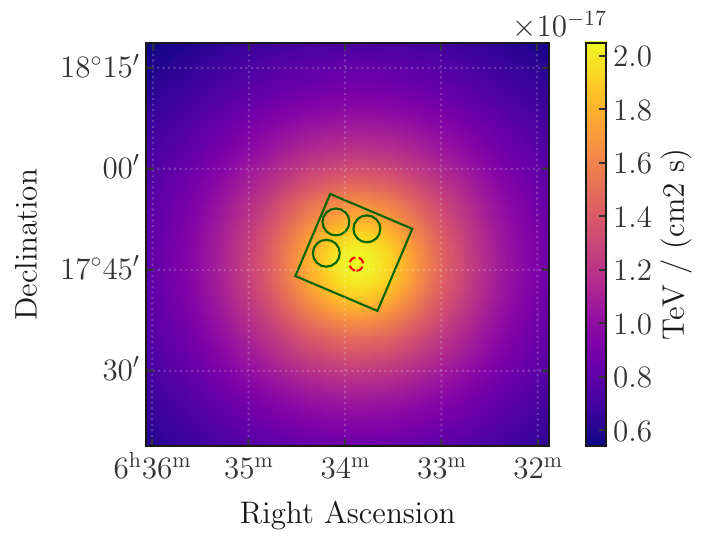}   
\caption{Model sky maps of the Geminga X-ray halo
flux coming from synchrotron emission as integrated in the 8--40~keV band, and
overlaid with the \nustar\ FoVs for the focused and stray-light X-ray photon components for a squared region of $10\times 10^{\circ}$ (left panel) and the central $1\times1^{\circ}$ (right panel). 
The central $13'\times13'$ green squared region around the Geminga pulsar (red circle) will be focused through the \nustar\ optics, while X-rays within $r\sim$ 1 -- 4$^{\circ}$ region around the target (unshaded area in the left panel, for module A; see also Fig.~\ref{fig:modB_straylight}) will be detected as stray-light photons. The circular regions 1, 2 and 3 as defined in section~\ref{sec:nustar_obs} are also shown in the right panel.}
    \label{fig:model_skymap}
\end{figure*}

\begin{figure}
    \centering
\includegraphics[width=\linewidth]{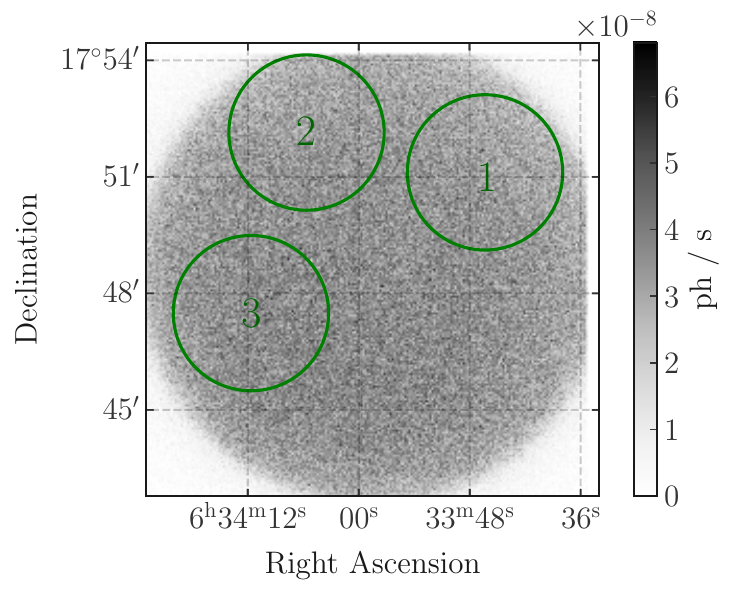}  
\caption{Count rate from the  \texttt{SIXTE} simulation of the Geminga synchrotron halo corresponding to the derived upper limits in the energy bin 8-40 keV. The emission from the Geminga pulsar is not included. The three circular regions used to extract the count rates to compare with the background subtracted data are also shown in green. The simulated exposure is $10^9$s.}
    \label{fig:sixte_skymap}
\end{figure}

Given that there was no X-ray halo detection in the \nustar\ data, we derive upper limits on the Geminga SED using the count rate upper limits. 
We used the physical Geminga synchrotron halo model, as described in \S \ref{sec:theory}, to simulate realistic count rates.
The general procedure are summarised in three steps: 
(i) computation of the physical model for the Geminga halo, (ii) simulation of the count rates for \nustar\, and (iii) comparison of the model count rates with the upper limits obtained using real data in the previous section. 
We perform these procedures iteratively for the energy bands $\Delta E =$ 8 - 40~keV and 40 -- 79~keV, over which the predictions and simulations are integrated. 
The normalisation of the model prediction increases gradually with the magnetic field value $B$ until the simulated count rate matches the observed upper limit value. When the upper limit is reached, we integrate the flux in a region of $\theta=0.23 ^{\circ}$ (\xmm{} FoV) to derive an upper limit on the SED. 
Given the substantial computational time of our innovative simulation procedure, we refrain from an extended investigation of the model dependency of the SED upper limits through the variation of the input model parameters. We nevertheless expect that the dependence would be very mild; what is controlling the final upper limits is the total flux emitted in a given energy bin and at a given angular scale. 

We begin by modelling the synchrotron surface brightness distribution $S(r, \Delta E)$ [photons\,cm$^{-2}$\,s$^{-1}$\,deg$^{-2}$] for a given energy band $\Delta E$. 
The benchmark model predictions used to simulate \nustar\ data are obtained using the parameters discussed in Section~\ref{sec:interpretation}, which match the existing Geminga halo observations from GeV to TeV energies. 
Starting from this benchmark model, we increase the value of the magnetic field $B$ until the simulated count rate matches the observed ULs. 
Figure \ref{fig:model_skymap} illustrates an example halo model map with the \nustar\ FoV for 2-bounce photons and for stray-light photons. 
In the left panel, we show the $10 \times 10^{\circ}$ region with the $13'\times13'$ green square around the Geminga pulsar corresponding to the \nustar\ observations. The sky regions contributing to stray-light photons for the \nustar{} FPMA are the unshaded region in the left panel, see appendix for the corresponding figure for the FPMB. The Geminga halo flux has been integrated in 8 -- 40~keV, and we show this quantity as computed for each pixel in units of TeV/cm$^2$/s.  
The right panel shows a zoomed-in version in the central $1\times1 ^{\circ}$, including the three circular regions selected for computing the upper limits. 

We then convolve the model surface brightness $S(r, \Delta E)$ with the \nustar\ optics (2-bounce photons) and collimator response functions (stray-light photons) separately. From each convolution process, we generated a count rate map on the detector plane in units of cts s$^{-1}$ pix$^{-1}$.

For 2-bounce X-ray halo photons originating from the central $13'\times13'$ region around the Geminga pulsar, we input the X-ray halo model map into the \texttt{SIXTE} simulation software package \citep{Dauser2019}. Since the optics effective area is energy dependent, we also fix the spectrum for each sky position grid in the form of $F(E) = N E^{-\Gamma}$. \texttt{SIXTE} is a generic X-ray telescope simulation tool that is applicable to simulating \nustar\ photon events. In each \texttt{SIXTE} simulation, we adopted $\sim 10^9$ s exposure time to ensure that statistical errors are negligible in the simulated data. Note that no background component is included in the simulation as we compare our model count rates with the background-subtracted count rates in Table \ref{tab:count_rates}. Also, the emission of the Geminga pulsar is not included in the simulation. 
An example of the simulated count rate (corresponding to the upper limit in the energy bin 8-40 keV, see next) is shown in Fig.~\ref{fig:sixte_skymap} in units of cts/s. We highlight the three regions used to extract the Geminga synchrotron halo count rate to be compared with the observational upper limits.

For stray-light X-ray halo photons from 1$^{\circ}$ -- $4^\circ$, we convolved a surface brightness model with the collimator response function constructed with the \texttt{NUSKYBGD} code \citep{2014Wik}, the approach widely used for CXB studies  \citep{Krivonos2021,2023AJ....166...20R}. Each detector pixel captures a different portion of the sky due to the optical bench shadowing and aperture stop.

Using our surface brightness model, we thus generated two separate count rate maps for 2-bounce and stray-light photons. These maps were then combined in each of the three circular regions, the contribution from stray-light emission summed to the \texttt{SIXTE} simulated one, and then compared to the net count rate upper limits derived in \S3.1. 
From our model convolution procedure, we find that the stray-light background component is as significant as the focused halo component, and dominant over the focused halo component in the 40-79~keV energy band. 
This demonstrates that to correctly interpret the upper limits on the count rates for an extended signal such as the one expected from the Geminga halo, both component needs to be taken into account. 

The sum of simulated count rates from 2-bounce and stray-light photons is used to translate the upper limit on the experimental count rates into an upper limit on the SED. We provide our results integrated within a region of $\theta=0.23^{\circ}$, matching the \xmm{} FoV. This is a convenient choice for aligning regions in the two instruments used for our analysis.
The outcome of the whole procedure is a flux upper limit of $2.2 \times 10^{-12}$ erg s$^{-1}$ cm$^{-2}$ for the 8 -- 40~keV band and of $3.5 \times 10^{-12}$ erg s$^{-1}$ cm$^{-2}$  for the 40 -- 79~keV band. 
Our results are shown with black points in Fig.~\ref{fig:xmm_interpr}, left panel, along with the upper limits obtained from \xmm{}. 
Altogether, we provide new upper bounds on the X-ray diffuse emission from the Geminga halo from 0.5 to 79 keV. 
Note that for \xmm{} we exclude the middle energy range [1.3, 2] keV  because of the strong background line emission, see section~\ref{sec:xmm_data}.
The  \nustar{} upper limits are found a bit lower that the adjacent limits from \xmm{}. However, as we will discuss in the next section, the lower energy band provides more constraints to the theoretical interpretation.

\section{Discussion}\label{sec:interpretation}
\label{sec:results}
\begin{figure*}
\centering
\includegraphics[width=0.49\textwidth]{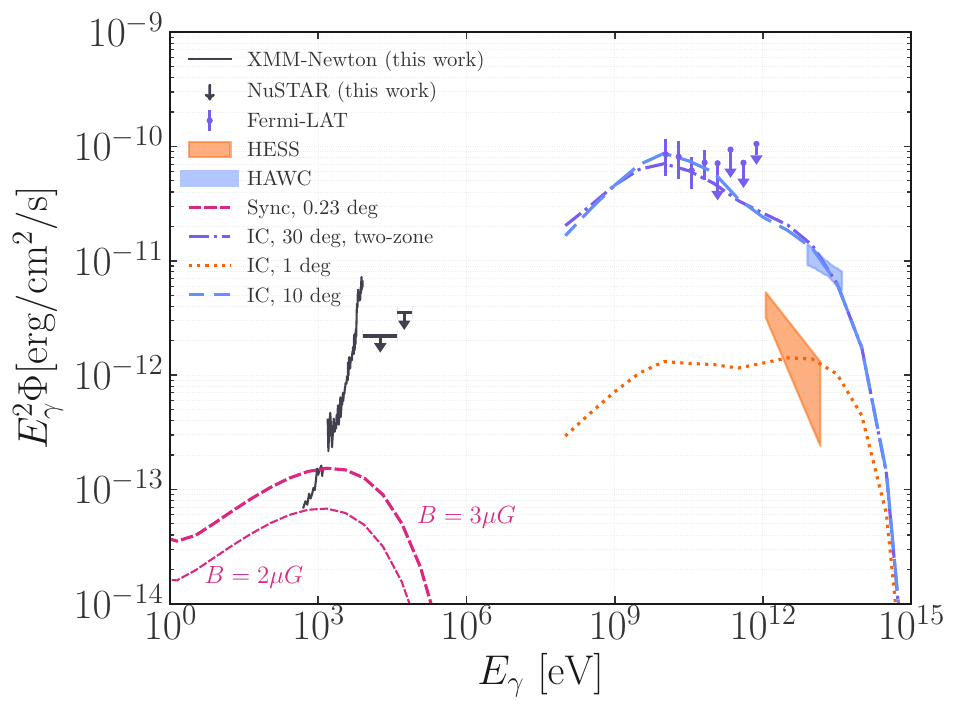}
\includegraphics[width=0.49\textwidth]{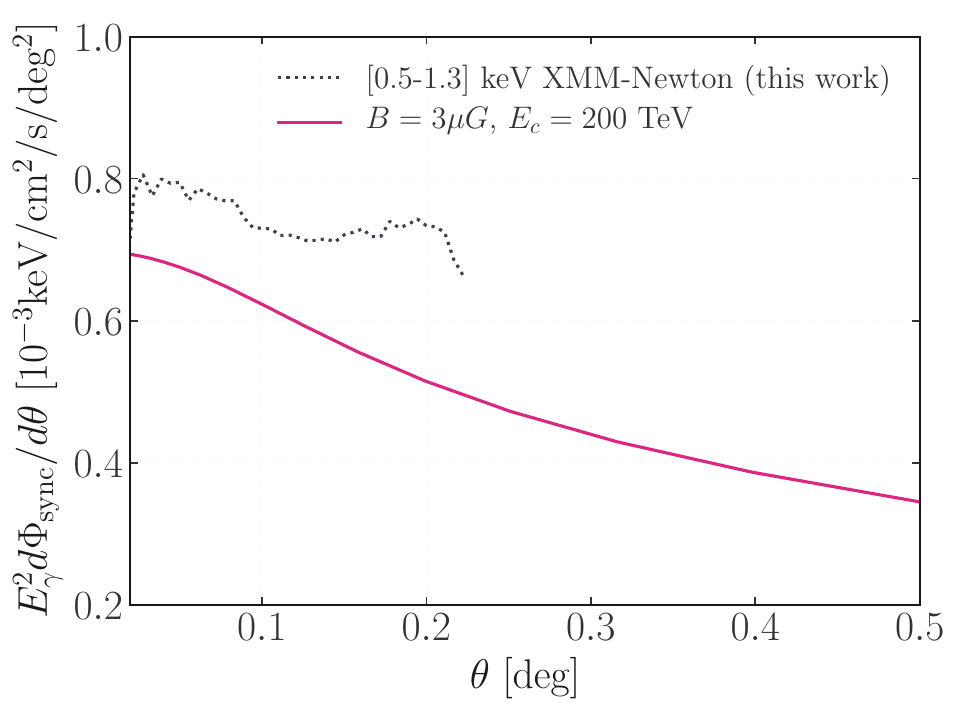}
\caption{\textit{Left panel}: interpretation of all available Geminga pulsar halo multiwavelength SED data, from the  gamma ray inverse Compton emission (\fermi{}-LAT, H.E.S.S., HAWC) to the X-ray upper limits (\xmm{}, \nustar{})  as obtained in this work. The inverse Compton and synchrotron emissions are obtained within a model in which electrons and positron pairs diffuse in a low diffusion zone, see text for details on the model parameters. 
\textit{Right panel}: comparison of the Geminga's synchrotron halo surface brightness   with the \xmm{} upper limits on the angular profile.  }
\label{fig:xmm_interpr}
\end{figure*}

In this section, we discuss the consequences of our theoretical model on the emission of X-rays around the Geminga pulsar. We reiterate that the main idea is that a population of energetic electrons accelerated by the pulsar emits X-ray photons by synchrotron emission. These same electrons are supposed to also up-scatter ambient photons to GeV and TeV energies by ICS.

We aim to combine our theoretical framework and the observational X-ray flux upper limits to constrain the model parameters. 
Since this framework is based on the hypothesis that the X-ray and the gamma-ray emissions point to the same population of electrons, we choose some free parameters so as to provide  
a reasonable explanation of the \fermi{}-LAT, HAWC and H.E.S.S. SED. 
Specifically, the relevant benchmark model parameters are
$\gamma=1.85$, $E_c=200$~TeV, $\eta=0.12$, a suppressed diffusion coefficient around Geminga of  $D_0=10^{26}$~cm$^2$s$^{-1}$, and a magnetic field of $B=3\mu$G, see section \ref{sec:theory} for their definition. 
 
In Fig.~\ref{fig:xmm_interpr}, we show the multiwavelength SED of the Geminga pulsar halo as obtained using these benchmark parameters. 
To consistently compare our predictions with the observations of the ICS emission of the Geminga pulsar halo, we integrate the emission model up to the angular scale relevant for each experiment ($\theta=1,10,30 ^{\circ}$ to for H.E.S.S., HAWC and \fermi{}-LAT, respectively). Each prediction is thus reported using a different line style in Fig.~\ref{fig:xmm_interpr}, but effectively corresponds to the same physical model. For a visual comparison, these predictions are extended from 0.1~GeV up to the multi-TeV energy range. 
A two-zone diffusion model is employed for comparing with \fermi{}-LAT data, fixing $r_b=90$~pc, as it has been demonstrated that at these energies and corresponding angular scales, it is better suited to model the observed flux. 
While we find good agreement with \fermi{}-LAT and HAWC data, our model does not completely reproduce the H.E.S.S. result at the lower energies. 
This is likely because the rescaled HAWC data do not perfectly match the H.E.S.S. result within $1^{\circ}$ \citep{HESS:2023sbf}. 
Indeed,  the HAWC and H.E.S.S. collaborations employed different assumptions for the  model to fit the data, i.e. its spatial morphology (diffusive vs. 2D disk), and they have different FoVs.  

Once we determined the electron injection spectrum and propagation properties on the GeV-TeV inverse Compton SED, we focus on interpreting the X-ray observation data.
The theoretical prediction for this emission mechanism is mostly ruled by the assumption of the strength of the magnetic field, as shown in section~\ref{sec:theory}. The value of B varies consistently both in the emission power and in the energy losses.

In the left panel of Fig.~\ref{fig:xmm_interpr} we display the predictions for the synchrotron flux considering  $B=2\mu$G and  $3\mu$G. 
The model emission is integrated in an angular scale of $\theta=0.23^{\circ}$ to approximately match the \xmm{} and \nustar{} FoV.
These predictions are compared with the new X-ray upper limits on the Geminga's halo flux obtained as described in section~\ref{sec:X-ray_data}. 
We find that the \xmm{} energy-dependent upper limits, and specifically the  0.5-1.3~keV energy range constrain the magnetic field within the Geminga's halo to be less than about $2 \mu$G, when fixing all the other parameters to the benchmark values. 
The \xmm{} upper limits strongly constrain the normalisation of the X-ray halo emission at around 1~keV. 
This result is the physical consequence of the assumption that the same population of electrons is responsible for the ICS and synchrotron emission. The energy shape on the X-ray emission is then unavoidably peaked at a few keV. 
We note that other parameter sets
could provide similar solutions or consistent results with the X-ray data, as the magnetic field is not the only parameter that determines the normalisation 
of the synchrotron contribution, as explored in section~\ref{sec:theory}. While we here explore the main dependencies,  a thorough investigation of the complete parameter space compatible with the multiwavelength  observations will be explored in the future work. 

The analysis of the \xmm{} data also delivers an upper limit on the surface brightness of Geminga's halo emission. These upper limits obtained using data between 0.5--1.3~keV are  shown in the right panel of Fig.~\ref{fig:xmm_interpr} together with the predicted surface brightness from the Geminga halo. 
One can easily see that the theoretical surface brightness for $B=3$ $\mu$G is compatible with the \xmm{} upper limits,  thus it is less effective for constraining the model parameters compared to using the SED data.  
The upper limits for the 2--8~keV surface brightness are found to be a factor $\sim 5$ higher with respect to the benchmark prediction, and are reported in the appendix for completeness. 

The upper limit we obtain for the magnetic field in a region of $0.23^{\circ}$ around Geminga is consistent with the one obtained in \cite{Khokhriakova:2023rqh}, which is found to be of 1.4$\mu$G when using {\it eRosita} data in the energy band 0.5--2~keV. 
Moreover, our constraint on $B$ is fully compatible with the value predicted by the magnetic field model of ~\cite{2012ApJ...761L..11J,2023arXiv231112120U}, which is of about 1.6 $\mu$G at the pulsar location.

\begin{figure*}
\centering
\includegraphics[width=0.49\linewidth]{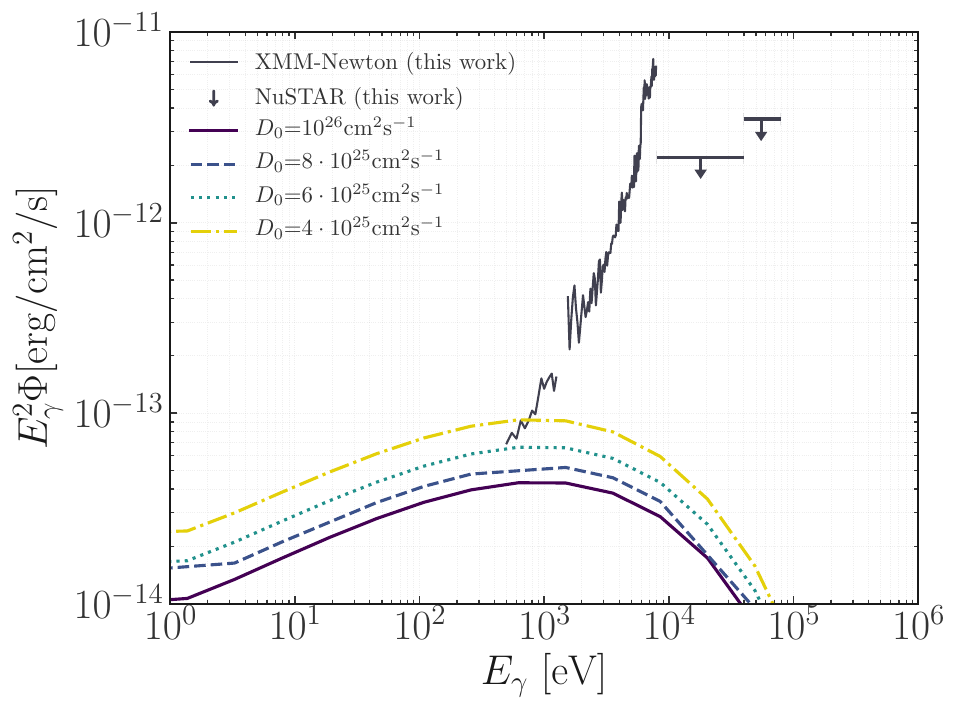}
\includegraphics[width=0.49\linewidth]{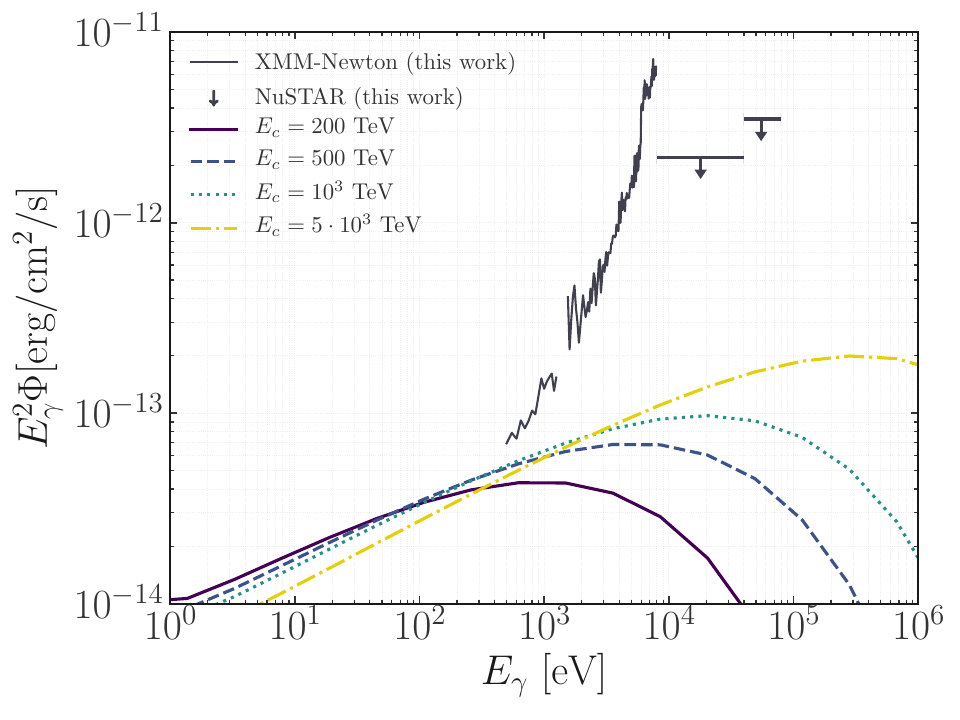}
\caption{The synchrotron emission from the Geminga pulsar halo as compared to the upper limits in the X-ray band obtained in this work using \xmm{} and \nustar{} data when varying the diffusion coefficient $D_0$ (left panel) and the energy cutoff of the electron source term (right panel). See text for details. }
\label{fig:D0_Ec_interpt}
\end{figure*}

As explored in section \ref{sec:theory}, the particle diffusion coefficient affects the X-ray halo normalisation and size and thus the predicted flux. Although the surface brightness of the TeV halo emission can strongly constrain the diffusion coefficient $D_0$, we demonstrate a complementary role  of our X-ray results regarding this crucial parameter in the left panel of Fig.~\ref{fig:D0_Ec_interpt}. 
By fixing the magnetic field around Geminga of $B=1.6\mu$G \citep{2012ApJ...761L..11J} and the other parameters to their benchmark values, 
the \xmm{} upper limits are reached for $D_0=6 \times 10^{25}$ cm$^2$ s$^{-1}$. Higher values are excluded by the stringent constraints at around 1~keV. 
This illustrates the complementary constraining power of the X-ray measurements in the keV energy band. 

Our X-ray constraints cover the previously unexplored energy range of $E_\gamma>10$~keV by means of the \nustar{} data. This energy range could be crucial  
to constrain the spectral shape of the synchrotron emission, as showcased in the right panel  of Fig.~\ref{fig:D0_Ec_interpt}. Within the benchmark parameters, and fixing $B=1.6\mu$G, we illustrate how increasing the cutoff energy $E_c$ of the electron source term significantly increases the synchrotron emission in the \nustar{} energy range. 
Although the current \nustar{} upper limits lie at about one order of magnitude below the explored configurations, we expect them to be  complementary to \xmm{} constraints for specific regions of the parameter space, e.g. for steeper injection spectra, see Fig.~\ref{fig:spectra_sb}.

\section{Conclusion}
\label{sec:conclusions}
In this paper, we present new results for the X-ray halo around the Geminga pulsar as extracted from  \xmm{} and \nustar{} archival data. 
The upper limits derived on the flux are interpreted phenomenologically in terms of a model taking into account suppressed diffusion and ICS and synchrotron energy losses of electrons and positrons accelerated by Geminga. 

We analyse the \xmm{} observations in the energy range of 0.5 -- 8 keV and in a field of view of 900" radius around the Geminga pulsar.   
Individual background components are derived using the control regions in which the background events are dominant.
This novel approach allows the measurement of the absolute level of Geminga halo emission in the signal region independent of the control regions, which is ideal for the source with an angular extension larger than the field of view of the instrument.
No significant extended emission is detected in our \xmm{} analysis. New X-ray flux limits are derived based on  the null detection of Geminga halo, and the systematic uncertainties on the individual background estimations are provided. They are found to span from  
 $7 \times 10^{-14}$ erg s$^{-1}$ cm$^{-2}$ to  $7 \times 10^{-12}$ erg s$^{-1}$ cm$^{-2}$ in the 0.5 -- 8 keV energy range. 

A \nustar\ archival data analysis in the interval 8 – 79~keV is performed here with a novel technique that considers both focused and stray-light X-ray components of spatially non-uniform diffuse X-ray emission. 
In particular,  we demonstrate the importance of stray-light background photons when searching for objects extending over a few degrees in the sky with \nustar{}.
Given the no X-ray halo detection in the \nustar\ data, we derive upper limits on the Geminga SED using a physical Geminga synchrotron halo model as a tool to simulate realistic counts rates.
We find a flux upper limit of $2.2 \times 10^{-12}$ erg s$^{-1}$ cm$^{-2}$ for the 8 -- 40~keV band and of $3.5 \times 10^{-12}$ erg s$^{-1}$ cm$^{-2}$ for the 40 -- 79~keV energy range. The \nustar\ flux limits are slightly lower than the adjacent limits from \xmm{}. 
Altogether, we provide new upper bounds on the X-ray diffuse emission from the Geminga halo from 0.5 to 79 keV. 

Our observational upper limits are then interpreted within a physical emission model based on the idea that a population of very energetic electrons accelerated by the pulsar emits X-ray photons by synchrotron emission $and$ up-scatters ambient photons to GeV and TeV energies by ICS. We therefore determined the electron energy and spatial distributions that match with the \fermi{}-LAT, HAWC and H.E.S.S. SED, and then focus our attention on the X-ray band. The theoretical prediction for this emission mechanism is mostly ruled by the assumption of the strength of the magnetic field. 
We find that the \xmm{} energy-dependent upper limits, and specifically the  0.5 -- 1.3~keV energy range, constrain the magnetic field within the Geminga's halo to be less than about $2 \mu$G, when fixing all the other parameters to the benchmark values. 
This result is the consequence of the physical assumption that the same population of electrons is responsible for the ICS and synchrotron emission. The energy shape on the X-ray emission is then unavoidably peaked at a few keV. 

Our constraint on $B$ is fully compatible with the value predicted by the magnetic field model of  ~\cite{2012ApJ...761L..11J, 2023arXiv231112120U}, which is of about 1.6 $\mu$G at the pulsar location. 
Fixing the magnetic field around Geminga to $B=1.6\mu$G  and the other parameters to their benchmark values, 
the \xmm{} upper limits are reached for $D_0=6 \times 10^{25}$ cm$^2$ s$^{-1}$. Higher values are excluded by the stringent constraints at around 1~keV. 
We also illustrate how increasing the cutoff energy $E_c$ of the electron source term significantly increases the synchrotron emission in the \nustar{} energy range. 

The first derivation of  upper limits of X-ray emission from two experiments and the interpretation of these upper limits in the context of a physical multi-wavelength model which traces back to very energetic electrons 
accelerated by the same source, is the major novelty of this paper. 
By combining the newly derived X-ray flux upper limits in different energy bands and regions of the Geminga halo, we present new results for constraining physical halo parameters using our physical multi-wavelength model.
Our results support the idea that 
broadband halo emission from keV to multi TeV around the Geminga pulsar originate by the same electron populations that are accelerated by Geminga itself. 
Our methodology can be applied to other pulsar halos observed by archival \nustar\ and \xmm{} data or future \nustar{}, \xmm{} or other X-ray observations.  In particular,
more distant TeV halos with smaller angular sizes could be more sensitively studied with \nustar{} because the focused X-ray component becomes more significant.
From our theoretical model, it is predicted that X-ray pulsar halos exist around pulsars, and future dedicated observations could reveal their existence around some of these objects. 
In the future, we plan to search for diffuse X-ray emission associated with other pulsar halos and leptonic Galactic PeVatron candidates with \nustar\ and \xmm{} observations, and help to provide deeper insights into the still mysterious origin of the gamma-ray halos shining around these objects.

\begin{acknowledgements} We thank Reshmi Mukherjee for her careful reading and insightful discussion of the manuscript.  
SM acknowledges the European Union's Horizon Europe research and innovation program for support under the Marie Sklodowska-Curie Action HE MSCA PF–2021,  grant agreement No.10106280, project \textit{VerSi}.
RYS acknowledges  NSF award PHY211049. CT acknowledges NuSTAR NASA Grant number 80NSSC22K0573. 
M.D.M. and  F.D. acknowledge the support of the Research grant {\sc TAsP} (Theoretical Astroparticle Physics) funded by Istituto Nazionale di Fisica Nucleare. 
F.D. acknowledges the Research grant {\sl Addressing systematic uncertainties in searches for dark matter}, Grant No. 2022F2843L funded by the Italian Ministry of Education, University and Research (MIUR).
\end{acknowledgements}

\bibliographystyle{aa} 
\bibliography{biblio}


\begin{appendix}
\section{Additional material}
In this section, we report a number of additional figures and information to complement what is illustrated in the main text, both for the X-ray analysis as well as for the model interpretation. 

\begin{figure}
\centering
\includegraphics[width=\linewidth]{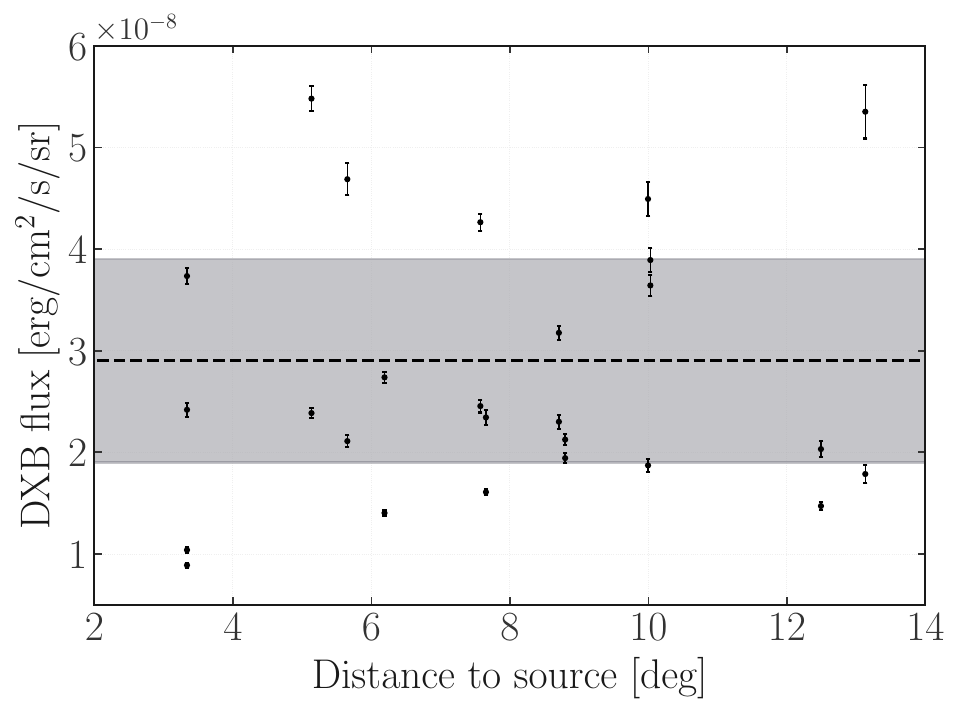}
\includegraphics[width=\linewidth]{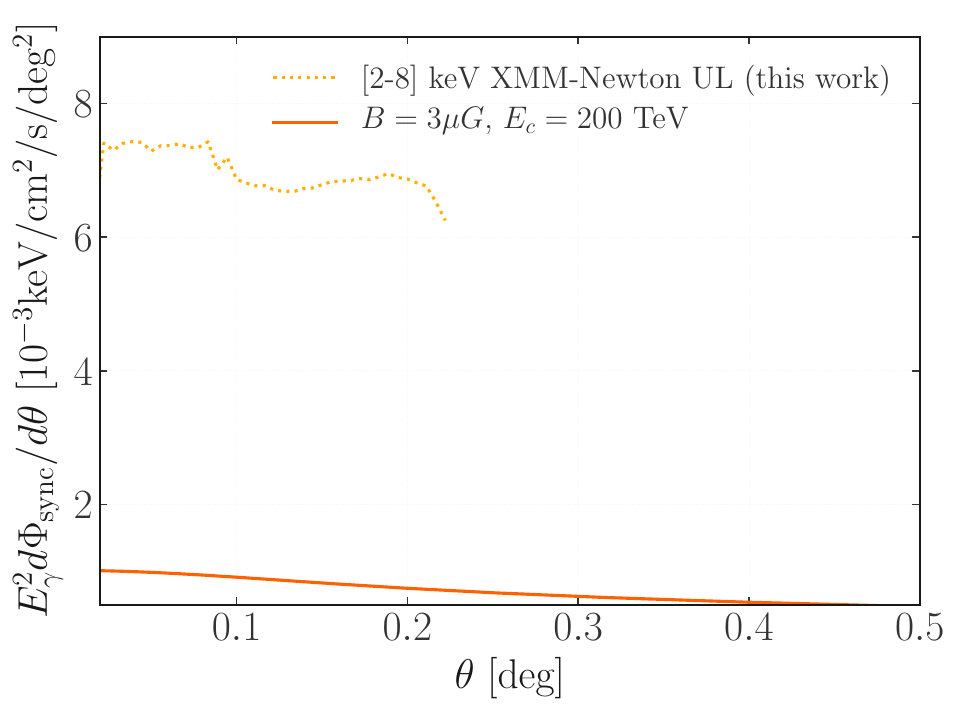}
\caption{ 
Upper panel: The measured DXB fluxes from multiple extragalactic observations with \xmm{}.
The vertical bars represent the statistical error in the measurement of each observation.
The grey band shows the averaged DXB flux and uncertainty, and the horizontal dashed line the average DXB flux. 
Lower panel: Comparison of Geminga model with \xmm{} upper limits on the angular profile for the high energy bin 2-8~keV.}
\label{fig:xmm}
\end{figure}

In the upper panel of Fig.~\ref{fig:xmm}, we show the measured DXB flux of each individual \xmm{} observations as a function of the angular distance to the Geminga pulsar (black points), together with the estimated averaged DXB flux and uncertainty (grey band). We note that the measured DXB flux from the Geminga ROI does not show any clear trend as a function of the distance, meaning that this component is not absorbing any residual X-ray halo emission. 

In the lower panel of Fig.~\ref{fig:xmm} we report the result for the angular profile upper limits as obtained in the high energy bin of the \xmm{} data analysis, 2-8~keV (yellow dotted line). This is compared with the prediction of the model tested by the SED \xmm{} upper limits shown in Fig.~\ref{fig:xmm_interpr}, corresponding to a $3\mu$G magnetic field. For this specific realization, the high energy surface brightness upper limits are found to be a factor of about 5 higher than the model prediction and thus do not add constraining power. 

\begin{figure*}
\centering
\includegraphics[width=0.55\linewidth]{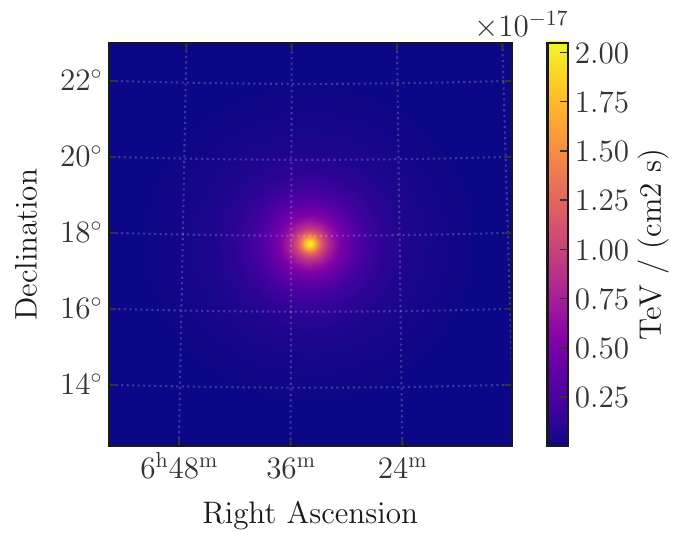}
\includegraphics[width=0.44\linewidth]{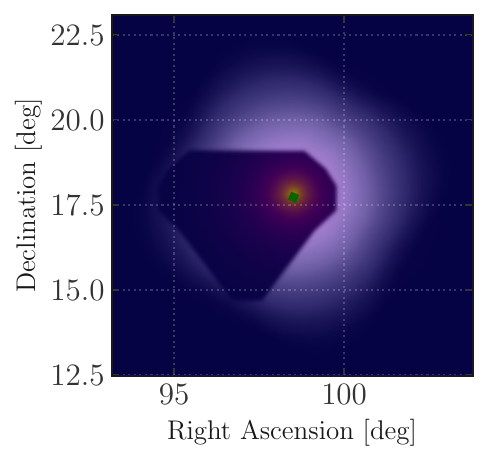}
\caption{ In the left panel, we show the model sky map of the Geminga X-ray halo flux coming from synchrotron emission as integrated in the energy bin 8--40~keV. In the right panel, we overlay the stray-light X-ray photon FoV to this model for \nustar{} module B.  This figure is complementary to Fig.~\ref{fig:model_skymap}, left panel.}
\label{fig:modB_straylight}
\end{figure*}

In Fig.~\ref{fig:modB_straylight}, we present a complementary view of what is shown in the main text in Fig.~\ref{fig:model_skymap}. The left panel shows the model sky map of the Geminga X-ray halo flux as in the left panel of Fig.~\ref{fig:model_skymap}, but without overlaying the stray-light \nustar{} FoV, to underline the spatial shape of the model at large angular scales. The right panel is the same as Fig.~\ref{fig:model_skymap} (left) but for \nustar{} FPMB observations. 

\end{appendix}

\end{document}